\pdfoutput=1
\documentclass[11pt]{article}
\usepackage{amsmath,amsfonts,mathrsfs,amssymb}
\usepackage{bm}
\usepackage{dsfont}
\usepackage{mathtools}
\usepackage[dvipsnames]{xcolor}
\usepackage{tensor}
\usepackage{slashed}

\usepackage{setspace}
\usepackage[font=footnotesize]{caption}
\usepackage{textcomp}

\usepackage{graphicx}
\setkeys{Gin}{draft=false}
\graphicspath{{./}{./Figures/}{./figures/}{./Figures/}}
\DeclareGraphicsExtensions{.pdf,.png,.jpg,.jpeg}

\usepackage[colorlinks]{hyperref}
\hypersetup{linktocpage}
\usepackage[titletoc]{appendix}
\usepackage{cite}
\definecolor{carmine}{rgb}{0.59, 0.0, 0.09}
\definecolor{sgreen}{rgb}{0.0, 0.44, 0.0}
\definecolor{prussianblue}{rgb}{0.0, 0.06, 0.54}
\hypersetup{colorlinks=true, linkcolor=carmine, filecolor=magenta, urlcolor=prussianblue, citecolor=sgreen}

\providecommand{\keywords}[1]{%
  \small
  \noindent\textbf{\textit{Keywords---}} #1\par
}

\definecolor{darkgreen}{rgb}{0,0.35,0}

\newcommand{\CECs}{Centro de Estudios Cient\'ificos (CECs), Casilla 1469, Valdivia, Chile}
\newcommand{\USS}{Facultad de Ingenier\'{\i}a, Universidad San Sebasti\'an, General Lagos 1163, Valdivia 5110693, Chile}
\newcommand{\USSs}{Facultad de Ingenier\'{\i}a, Universidad San Sebasti\'an, Bellavista 7, Santiago 8420524, Chile}
\newcommand{\UDP}{Instituto de Ciencias B\'asicas, Facultad de Ingenier\'{\i}a y Ciencias, Universidad Diego Portales, Ej\'ercito 441, Santiago 8370191, Chile}

\begin{document}

\setlength\arraycolsep{2pt}

\renewcommand{\theequation}{\thesection.\arabic{equation}}
\makeatletter\@addtoreset{equation}{section}\makeatother
\setcounter{page}{1}

\hypersetup{pageanchor=false}
\begin{titlepage}
\begin{center}

\vskip 1.0 cm

\begin{doublespace}
{\huge\bf Chiral--Maxwell Cavity EFT:\\ Reduced Gauge-Mode Condensation and Quantum-Optics Limits}
\end{doublespace}

\vskip 1.6cm

{\Large 
Fabrizio Canfora$^{1,2}$, Mauricio Ipinza$^{3}$, and Sim\'on Riquelme$^{4}$
}

\vskip 0.5cm

{\it 
$^{1}$\CECs\\
$^{2}$\USS\\
$^{3}$\UDP\\
$^{4}$\USSs
}

\vskip 0.4cm

{\small
\texttt{fabrizio.canfora@uss.cl},\quad
\texttt{mauricio.ipinza@mail.udp.cl},\quad
\texttt{sriquelm@ing.uchile.cl}$^{\,*}$
}

\vskip 0.15cm
{\small $^{*}$Corresponding author}

\vskip 0.2cm
{\small Accepted in \emph{The European Physical Journal C}. DOI: \texttt{10.1140/epjc/s10052-026-16250-6}}

\vskip 1.4cm

\end{center}

\vspace{-1.2cm}
\begin{abstract}
We develop a finite-volume Chiral--Maxwell effective field theory in which a topologically nontrivial hadronic sector acts as a nonlinear medium for a reduced cavity gauge mode. Starting from the \(SU(2)\) nonlinear sigma model minimally coupled to Maxwell theory, we construct an equivariant low-mode sector that retains the topological charge structure while projecting the dynamics onto an effective \(1+1\) theory. Both mass-like scales in the reduced action are finite-volume kinematic gaps---arising from transverse winding energy and holonomy curvature, respectively---with no explicit symmetry breaking, in direct analogy with the Hosotani mechanism on compact cycles. Both gaps vanish in the infinite-volume limit, consistently with massless photons and pions in bulk ChPT. In one hierarchy, Coleman duality yields a gauge-invariant one-loop effective potential for the reduced gauge coordinate; we identify its domain of validity and establish that the locally stable nontrivial branch satisfies the EFT hierarchy condition. In the complementary hierarchy, integrating out the gauge mode produces a one-loop deformed sine--Gordon EFT for the chiral sector with a closed-form kink tension correction. Upon quantization, the reduced theory maps onto nonlinear cavity Hamiltonians whose trivial and locally condensed branches obey distinct selection rules, providing experimentally accessible diagnostics.
\end{abstract}

\noindent
\begin{minipage}{\textwidth}
\keywords{Chiral Perturbation Theory; finite-volume effective field theory; Hosotani mechanism; \mbox{sine-Gordon} model; topological solitons; cavity quantum electrodynamics}
\end{minipage}

\end{titlepage}

\hypersetup{pageanchor=true}

\newpage

\tableofcontents

\newpage

\section{Introduction}
Spatially modulated phases are a recurrent theme in finite-density physics: once a conserved charge is forced into a finite region, ordering at nonzero momentum can become energetically favorable and translational invariance may be broken by an inhomogeneous condensate. In QCD at finite baryon density and low temperature, related expectations are often phrased in LOFF-type terms \cite{R1,R2,newd5,newd6,FuldeFerrell1964,LarkinOvchinnikov1965}. In the low-energy regime, QCD coupled to electromagnetism is described by Chiral Perturbation Theory (ChPT) minimally coupled to Maxwell theory \cite{ChPT1,ChPT2,ChPT3,ChPT4,ChPT5,ChPT6,ChPT7,ChPT8}. In this framework baryon number is encoded as a topological charge \cite{skyrme,Witten,witten0,bala0,Bala1PRL,Bala1NPB}, so finite-volume and topological constraints naturally promote spatial structure. In solvable \(1+1\) settings, finite-density regions can be dominated by kink crystals \cite{toy1,toy2,newd13y16a,newd13y16b,newd13y16c,newd13y16d,newd7,newd20toN3}, while in nuclear-matter contexts forcing baryonic charge into a finite volume leads to ``pasta''-like phases \cite{pasta1}. These structured hadronic backgrounds motivate a concrete question: how does electromagnetic radiation propagate through, and backreact on, such media?

Analytic progress benefits from controlled geometries where the relevant low modes can be isolated explicitly. The finite-volume technology developed in \cite{Fab0,Fab1,gaugsk,gaugsk2,crystal1,crystal2,crystal3,range1,range2,range3,range4} provides exactly this: it identifies parametrizations of the \(SU(2)\)-valued chiral field that capture inhomogeneous, topologically nontrivial configurations in cavity- and waveguide-like domains, while keeping the dynamics analytically tractable. This setting is therefore well suited to address strong-coupling ``light-matter'' interactions in hadronic media. The precise mixed Chiral--Maxwell cavity sector studied here appears to be new, but its logic belongs to this broader class of finite-volume and equivariant constructions: topology is preserved by allowing nontrivial winding along compact directions, while the remaining slow dynamics is projected onto the longitudinal low modes. The value of this approach lies precisely in analytical tractability: much as the Schwinger model and the sine--Gordon theory illuminate confinement and topological charge in settings more amenable to exact analysis than four-dimensional QCD, the reduced sector developed here provides a controlled environment where the interplay between gauge holonomy and hadronic topology can be followed exactly through the one-loop level.

The rectangular compactification used below should therefore not be understood as a unique experimental geometry, but as a controlled representative of a broader class of finite-volume equivariant sectors. Compact toroidal and cylindrical geometries of exactly this type have appeared before in several distinct contexts: in finite-density hadronic constructions and nuclear-pasta modeling, where forcing baryonic charge into a compact domain selects periodic or waveguide-like spatial structures \cite{pasta1,Fab0,Fab1}; in analyses of gauge theories on small circles, where holonomy (Wilson-line) zero-modes become light and can condense through the Hosotani mechanism \cite{Hosotani83}; and in equivariant Skyrme and Chiral--Maxwell reductions designed to retain topological sectors while keeping the dynamics analytically tractable \cite{gaugsk,gaugsk2,crystal1,crystal2,crystal3,CanforaMaeda2013,CanforaDiMauroKurkovNaddeo2015}. Our construction belongs to this established family. Other compact transverse manifolds or boundary conditions lead to analogous reductions, with modified spectra, mode overlaps, and numerical coefficients, provided that the transverse winding data and the longitudinal low modes remain separable.

A complementary motivation comes from cavity quantum electrodynamics and photon condensation in optics and condensed matter. Cavity QED is a mature framework for studying how spatial confinement and resonant boundary conditions reshape light--matter interactions, from the modification of spontaneous emission to the strong-coupling regime of single atoms, collective atomic gases, and many-body quantum gases \cite{HarocheKleppner1989,Walther2006CQED,Kimble1998CQED,Brennecke2007BEC,Mivehvar2021QuantumGases}. The same confinement logic makes the status of gauge-invariant cavity variables especially sharp: once light--matter coupling becomes ultrastrong, truncations that are not carefully tied to the underlying gauge structure can spoil spectra and correlation functions \cite{DiStefano2019Gauge,Garziano2020Gauge,Settineri2021Gauge,Savasta2021Gauge,Salmon2022Gauge,Akbari2023Gauge,Hughes2024Gauge}. In blackbody radiation the photon chemical potential vanishes, so photon number is not independently tunable and photons do not condense. Achieving photon condensation requires both thermalization mechanisms and an effectively conserved photon number, typically realized under strong light--matter coupling in suitable cavities. Early theoretical proposals include thermalization via Compton scattering in plasmas or nonlinear resonators \cite{photoncond,photoncond0}, while the strong-coupling regime and mixed light--matter states provide a more favorable arena \cite{photoncond0b,photoncond0c,photoncond0d}. Direct observations of photon condensation have since been reported in several platforms \cite{photoncond1,photoncond2,photoncond3,photoncond4,photoncond5,photoncond6}. Related geometric formulations of Maxwell theory, organized around helicity, electromagnetic duality, and optical chirality, reinforce the same lesson from a different angle: global field invariants can become the natural degrees of freedom once confinement and polarization structure are taken seriously \cite{Bliokh2011OpticalChirality,Bliokh2013DualEM,Bliokh2014Conservation,Bliokh2015AngularMomenta,Dressel2015SpacetimeAlgebra,Alpeggiani2018Helicity,Bliokh2019SurfaceMaxwell,Bliokh2019Acoustic}. Our aim is not to reproduce any particular atomic, polaritonic, or quantum-gas implementation. Rather, we use the logic of cavity-enhanced light--matter coupling to extend the cavity-QED viewpoint to hadronic/topological degrees of freedom and to ask whether they can generate, within a controlled finite-volume field theory, a locally condensed branch for a reduced cavity gauge mode.

We work in a finite-volume cavity where the natural low-energy gauge-sector variable is a gauge-invariant global mode, i.e. Wilson-line/holonomy data on the compact cycles rather than a local expectation value of $A_\mu$ \cite{Hosotani83,AitkenChermanPoppitzYaffe2017,Akamatsu2022}. In this respect our approach shares the cavity-QED insistence on gauge-invariant observables \cite{DiStefano2019Gauge,Settineri2021Gauge}, while selecting those invariants from the compact geometry itself. It should be read as a finite-volume topological sector of the Chiral--Maxwell theory, not as the infinite-volume vacuum of ordinary low-energy QCD coupled to electromagnetism. This distinction is essential: a homogeneous reduction in which all chiral coordinates depended only on \((t,x)\) would erase the topological density, whereas the equivariant winding along the compact transverse directions keeps the reduced sector topologically nontrivial while leaving a controlled \(1+1\) local dynamics. We then build a consistent low-mode truncation of the coupled Chiral--Maxwell system that retains the local rank structure needed for nontrivial finite-volume baryonic/topological sectors.

The reduced dynamics is an effective \(1+1\) theory with two controlled hierarchies. In one hierarchy, integrating out the heavy hadronic sector produces a one-loop potential for the reduced gauge coordinate and an analytic local-condensation window; in the opposite hierarchy, integrating out the heavy gauge mode yields a one-loop corrected sine--Gordon EFT for the chiral sector. The characteristic scales that appear after the reduction are finite-volume gaps of the retained coordinates rather than ordinary particle masses in infinite-volume ChPT; their detailed interpretation follows from the reduced action itself.

By a reduced gauge-mode branch we mean, more precisely, a nontrivial local minimum of the effective potential for the lowest gauge coordinate within the reduced EFT. Its interpretation is restricted to the regime where the corresponding one-loop hierarchy is reliable; the resulting potential is not used as a global large-field completion of the full \(3+1\) theory. Upon quantization, the same reduction maps onto standard nonlinear quantum-optics Hamiltonians whose selection rules sharply distinguish the trivial and locally condensed branches, providing operator-level diagnostics of the reduced branches after an additional single-spatial-mode truncation.

Throughout we work at leading order in ChPT, i.e. the \(SU(2)\) nonlinear sigma model minimally coupled to Maxwell theory. This already captures the effects we want to isolate, so we postpone subleading operators like Skyrme, Wess--Zumino--Witten (WZW), higher-derivative terms, and related corrections to future work. We will, however, be explicit about the parametric regimes where integrating out either sector is reliable, and about the limitations imposed by finite volume.

To fix notation and establish the object whose topology must be preserved by the equivariant reduction, we recall the starting point. The action for a \(SU(2)\) chiral model coupled to electromagnetism is
\begin{align}
S[U,A]\ &= \ \, \int_{\mathcal{M}}\sqrt{-g}\ d^{4}x \,\left\{\frac{K}{4} \, \text{Tr}%
\left( R_{\mu }R^{\mu }\right) - \frac{1}{4\,e^2}\tensor{F}{_\mu_\nu}\tensor{F}{^\mu^\nu} \right\}\;,\label{eq:ChiralMaxwellAction}\\[0.35em]
R_{\mu } &\equiv U^{-1}D_{\mu}U = R_{\mu}^{a}\,t_{a},\quad D_\mu U \equiv \partial_\mu U + A_\mu [t_3,U] \;, \notag
\end{align}
where $U(x)\in SU(2)$, $D_\mu$ is the covariant derivative associated to the $U(1)$ gauge field $A_\mu$, whose gauge coupling is denoted by $e$, $\tensor{F}{_\mu_\nu} \equiv \partial_\mu A_\nu - \partial_\nu A_\mu$, and $t_{a}=i\sigma _{a}$ are the generators of the $SU(2)$ Lie group with $\sigma _{a}$ the usual Pauli matrices. With this convention one has, in particular, $\text{Tr}(t_a t_b) = -2\,\delta_{ab}$. The space-time manifold is split as $\mathcal{M} \simeq \mathds{R}\times\Sigma$, and $K$ is a dimensionful coupling of mass dimension $2$ which must be experimentally fixed.
The equations of motion for this system are given by
\begin{align}
D_\mu R^\mu = 0 \qquad \text{and} \qquad
\partial_\mu\tensor{F}{^\mu^\nu} = J^\nu,
\end{align}
where $J^\mu \equiv -\dfrac{e^2K}{2}\text{Tr}\left\{U^{-1}[t_3,U]R^\mu\right\}$ is the electromagnetic current of the model.

The topological current $\mathfrak{J}^\mu$ and the baryonic charge $\mathcal{Q}_{B}$ are defined as in \cite{skyrme,witten0,Witten},
\begin{align}
\mathfrak{J}^\mu\ \equiv \tensor{\epsilon}{^\mu^\nu^\alpha^\beta}\,
\text{Tr}\!\left[\left(U^{-1}\partial_{\nu}U\right)\left(U^{-1}\partial_{\alpha}U\right)\left(U^{-1}\partial_{\beta}U\right)\right] - 3\,\tensor{\epsilon}{^\mu^\nu^\alpha^\beta}\,\partial_\nu\,
\text{Tr}\!\left[A_\alpha t_3\left\{U^{-1}\partial_\beta U + \left(\partial_\beta U\right)U^{-1}\right\}\right],\label{currents}
\end{align}
and
\begin{align}
\mathcal{Q}_{B}\ \equiv \frac{1}{24\pi ^{2}}\int_{\Sigma }\mathfrak{J}^{0},
\end{align}
respectively. We use the totally antisymmetric Levi--Civita tensor $\tensor{\epsilon}{^\mu^\nu^\alpha^\beta}$ normalized as ${\tensor{\epsilon}{^{0123}}=1/\sqrt{-g}}$.
The first term in \eqref{currents} is the standard winding-number current, while the second term is an exact contribution that provides the gauge-invariant completion in the presence of $A_\mu$. This definition is kinematical and does not require adding a WZW term to the action. In the present reduction, the exact piece enters only through boundary/global data and does not alter the local bulk dynamics within the truncated sector.

Geometrically, the integrand in \eqref{currents} is (up to normalization and an exact piece) the pullback of the volume form on $SU(2)\simeq S^{3}$.
Consequently, $\mathfrak{J}^\mu$ vanishes whenever the map $U$ has local rank smaller than three, i.e. when $U$ effectively depends on fewer than three independent coordinate combinations.
For instance, if two would-be independent degrees of freedom of $U$ share the same spatial dependence, then $\mathfrak{J}^{\mu}$ vanishes identically. This local-rank condition directly motivates the equivariant Ansatz of Sec.~\ref{Sec:Ansatz}: the transverse winding along the compact directions is chosen precisely to keep $\mathfrak{J}^\mu$ nonzero within the reduced sector.

The remainder of the manuscript proceeds in stages. Section~\ref{Sec:Ansatz} sets up the cavity geometry and the low-mode chiral/gauge Ans\"atze. Section~\ref{Sec:ECM} derives the corresponding effective \(1+1\) Chiral--Maxwell theory, its classical dynamics, and the Coleman-dual one-loop effective action for the reduced gauge coordinate, including the local-condensation window. Section~\ref{Sec:EFTChi} develops the complementary one-loop chiral sine--Gordon EFT obtained by integrating out the reduced gauge mode. Section~\ref{Sec:QO} extracts the quantum-optics Hamiltonians and diagnostics. Section~\ref{Sec:Conclusions} collects discussion and outlook. Appendix~\ref{App:LieAnsatz} spells out the equivariant (Lie-derivative) rationale behind the reduction. With the general setup fixed, we turn to the cavity geometry and to the mode-truncation Ans\"atze that define the low-energy sector studied in the rest of the paper.

\newpage

\section{Ansatz}\label{Sec:Ansatz}

As discussed in the Introduction, our goal is to isolate a finite-volume Chiral--Maxwell sector in which a topologically nontrivial hadronic configuration couples strongly to a reduced cavity gauge mode. We therefore consider a ``box'' geometry with spatial volume $V=8\pi^{2}L_{x}L^{2}$ and metric
\begin{equation}
ds^{2}=-dt^{2} + dx^{2}+L^{2}(d \mathfrak{y} ^{2}+d \mathfrak{z} ^{2}) \;,
\end{equation}
where $L_x$ and $L$ set, respectively, the longitudinal and transverse scales. The coordinates range as
\begin{equation}  
0\leq x\leq L_x \,, \quad \  0\leq \mathfrak{y} \leq 2\pi \,, \quad 
0\leq \mathfrak{z} \leq 4\pi \;.
\end{equation}
Here $x$ has dimensions of length, whereas $\mathfrak{y}$ and $\mathfrak{z}$ are dimensionless. This keeps the interplay between $L$ and $L_x$ transparent in the reduction to an effective \(1+1\) description. In the limit $L_x\gg L$ the cavity becomes a \textquotedblleft nuclear wire\textquotedblright\ (or waveguide), with $(\mathfrak{y},\mathfrak{z})$ acting as transverse coordinates. The choice of a rectangular torus $T^2$ for the transverse section is not fine-tuned: any compact transverse surface admitting an equivariant action of the internal $U(1)$ with a nonzero winding integer $p$ leads to the same qualitative structure, with the specific geometry affecting only the numerical values of the characteristic mass gaps and the overall normalization of the reduced action.

A convenient parametrization of the $SU(2)$-valued chiral field is provided by Euler angles,
\begin{align}
U(x^\mu) = \exp \left( t_{3}\,F(x^\mu)\right) \exp \left( t_{2}\,H(x^\mu)\right) \exp \left(
t_{3}\,G(x^\mu)\right),
\label{EulerAnsatz}
\end{align}
where $F=F(x^\mu)$, $G=G(x^\mu)$ and $H=H(x^\mu)$ are the three scalar degrees of freedom of $U$. In this parametrization $H$ is traditionally referred to as the \textit{profile}.

To obtain an effective \(1+1\) dynamics while retaining the local rank structure required for nontrivial baryonic/topological sectors, one cannot simply demand that the three chiral degrees of freedom depend only on \((t,x)\): then the pullback entering \eqref{currents} has local rank at most two and the topological density vanishes. The way around is an equivariant (symmetry-up-to-gauge) construction: the fields wind rigidly along the compact transverse cycles, yet all gauge-invariant local densities (and in particular the energy--momentum tensor) remain independent of \((\mathfrak y,\mathfrak z)\). This strategy underlies the finite-volume constructions in \cite{Fab0,Fab1,gaugsk,gaugsk2,crystal1,crystal2,crystal3,range1,range2,range3,range4}. With this in mind, the $SU(2)$ Ansatz we adopt is
\begin{align}
H(x^\mu) &= H(t,x),\quad F(x^\mu) = p\,\mathfrak{z},\quad G(x^\mu) = p\,\mathfrak{y},
\label{Ansatz01}
\end{align}%
where we take $p\in\mathbb{Z}$ so that $U$ is single-valued under $\mathfrak{y}\to \mathfrak{y}+2\pi$ and $\mathfrak{z}\to \mathfrak{z}+4\pi$. This Ansatz has three key features. First, it reduces the three coupled nonlinear chiral equations to a single partial differential equation for the profile $H(t,x)$, which takes a sine--Gordon-like form in \(1+1\) dimensions. Second, despite the reduced dynamics, the rigid winding along the compact directions keeps the configuration genuinely three-dimensional in the local topological sense and can support arbitrarily large reduced baryonic charge once appropriate boundary/global data are selected. Third, for energy/temperature scales below $1/L$, the only relevant chiral fluctuations are $\delta H(t,x)$: any fluctuation with nontrivial dependence on $(\mathfrak{y},\mathfrak{z})$ is gapped by an energy of order $1/L$ and can be consistently neglected in a low-energy effective description.

The electromagnetic sector is chosen in the same spirit. At energies $\ll 1/L$ the relevant gauge dynamics is captured by the lowest mode compatible with the cavity symmetries; we parametrize this retained gauge coordinate by $\psi(t,x)$ and take
\begin{align}
A_\mu(x^\nu) &= \left(0,0,\frac{1}{2}\left(p - L\,\psi(t,x)\right),-\frac{1}{2}\left(p - L\,\psi(t,x)\right)\right).
\label{AnsatzGauge}
\end{align}%
The constant pieces along the compact directions encode the allowed global data in the cavity and are tied to the chiral winding labeled by $p$ through the covariant derivative. With \eqref{Ansatz01} and \eqref{AnsatzGauge}, local gauge-invariant quantities entering the action become independent of $(\mathfrak{y},\mathfrak{z})$, and the coupled Chiral--Maxwell equations close consistently within the $(t,x)$ sector. As discussed in Appendix~\ref{App:LieAnsatz}, this ansatz describes the low-energy mode in the cavity.

Substituting \eqref{Ansatz01} and \eqref{AnsatzGauge} into the full action and integrating over $(\mathfrak{y},\mathfrak{z})$ yields a reduced \(1+1\) effective action. Its Euler--Lagrange equations coincide with the restriction of the full $3+1$ equations to the Ansatz sector, so the truncation is consistent. Topologically, however, the configuration remains genuinely three-dimensional: the rigid windings in $(\mathfrak{y},\mathfrak{z})$ provide two independent spatial gradients, while the profile $H(t,x)$ supplies the third along $x$, so the gauge-invariant baryonic current \eqref{currents} need not vanish. For a Lie-derivative (equivariant) rationale behind the construction of the chiral and gauge Ans\"atze in \eqref{Ansatz01} and \eqref{AnsatzGauge}, including the role of compensated symmetries and Wilson-line zero-modes on compact cycles, see Appendix~\ref{App:LieAnsatz}.

\section{Effective Chiral--Maxwell Theory}\label{Sec:ECM}

Substituting the Ans\"atze of Sec.~\ref{Sec:Ansatz} into the action \eqref{eq:ChiralMaxwellAction} and integrating over the transverse coordinates $(\mathfrak{y},\mathfrak{z})$, we obtain a \(1+1\)-dimensional effective theory for the longitudinal fields. This section analyses that theory in two steps. First, we study the classical dynamics and identify the finite-volume mass-like scales that parametrize the spectrum (Sec.~\ref{subsec:classical}). Second, we integrate out the heavy hadronic sector via a Coleman-dual fermion and compute the one-loop effective potential for the reduced gauge coordinate, identifying the local-condensation window and its domain of validity (Secs.~\ref{subsec:Coleman}--\ref{subsec:validity}).

\subsection{Classical Perturbation Theory}\label{subsec:classical}

The reduced \(1+1\) effective Lagrangian is obtained by substituting the Ans\"atze \eqref{Ansatz01} and \eqref{AnsatzGauge} into the full \(3+1\) action and integrating over \((\mathfrak y,\mathfrak z)\), yielding
\begin{align}
\mathscr{L}\big[\varphi,\Psi\big]
&=
-\frac{1}{2}\,\partial_\mu\varphi\,\partial^\mu\varphi
-\frac{1}{2}\,\partial_\mu\Psi\,\partial^\mu\Psi
-2e^2K\,\Psi^2\sin^2\!\left(\frac{\beta\varphi}{2}\right)
-8Kp^2\pi^2\cos^2\!\left(\frac{\beta\varphi}{2}\right),
\end{align}
where we have introduced canonical fields adapted to the \(1+1\)–dimensional description,
\begin{align}
\varphi &\equiv \frac{2H}{\beta},
&
\beta &\equiv \frac{1}{\sqrt{2K}\,L\pi},
&
\Psi &\equiv \frac{2L\pi}{e}\,\psi.
\end{align}

We use the \(1+1\) Minkowski metric $\eta_{\mu\nu}=\mathrm{diag}(-1,+1)$, so $\square\equiv\partial_{\mu}\partial^{\mu}=-\partial_t^2+\partial_x^2$. Primes will denote $x$--derivatives.

At fixed time, in the reduced \(1+1\) description we define the (transverse-integrated) baryon density as
\begin{align}
\rho_B(x,t)
&\equiv \int d\mathfrak{y}\,d\mathfrak{z}\,\sqrt{\gamma_\perp}\ \mathfrak{J}^0(x,t),
\end{align}
with \(\mathfrak{J}^\mu\) given by \eqref{currents} and \(\gamma_\perp\) the induced metric in the \((\mathfrak{y},\mathfrak{z})\) section. In terms of the effective fields this becomes
\begin{align}
\rho_B(x,t)
&=
\frac{6pe}{\pi}\left[
\frac{\beta}{2}\,\sin\big(\beta\varphi(x,t)\big)\,\varphi'(x,t)\,\Psi(x,t)
-
\cos^2\!\left(\frac{\beta\varphi(x,t)}{2}\right)\Psi'(x,t)
\right]\nonumber\\[4pt]
&=
-\frac{6pe}{\pi}\,\frac{d}{dx}
\left[
\cos^2\!\left(\frac{\beta\varphi(x,t)}{2}\right)\Psi(x,t)
\right].
\end{align}
The last expression makes it explicit that
\(\rho_B\) is a total derivative in \(x\), so that at fixed time \(t\) the baryon
number reduces to a boundary contribution,
\begin{align}
B(t)
=
\int_{0}^{L_x}\!dx\,\rho_B(x,t) =
-\frac{6pe}{\pi}
\left[
\cos^2\!\left(\frac{\beta\varphi(x,t)}{2}\right)\Psi(x,t)
\right]_{x=0}^{x=L_x}.
\end{align}
The corresponding baryonic charge in the reduced \(1+1\) description is
\begin{align}
\mathcal{Q}_B(t)
&=
\frac{1}{24\pi^2}\,B(t).
\end{align}

The fact that the transverse-integrated density is a total derivative is not a contradiction with the nontrivial equivariant structure of the ansatz. The local rank-three structure is supplied by the two rigid transverse windings together with the longitudinal profile, while the gauge-invariant completion of the baryon current converts the reduced charge into a boundary/global datum. Thus a nonzero \(\mathcal{Q}_B\) in the reduced theory is selected by the allowed boundary values of the holonomy coordinate and of the chiral profile. In particular, the sector with \(\Psi=0\) and identical boundary data carries zero reduced charge, whereas nontrivial finite-volume topological sectors correspond to inequivalent boundary/global data of the pair \((U,A)\).

The classical equations of motion derived from the effective Lagrangian, which coincide with the complete field equations of the gauged nonlinear sigma model minimally coupled to electrodynamics once the reduction ansatz of the previous section is imposed, read
\begin{align}
\square\varphi
- e^2 K \beta\,\Psi^2\sin\!\big(\beta\varphi\big)
+ 4Kp^2\pi^2\beta\,\sin\!\big(\beta\varphi\big)
&= 0,\\
\square\Psi
- 4e^2 K\,\sin^2\!\left(\frac{\beta\varphi}{2}\right)\Psi
&= 0.
\end{align}
Setting \(\Psi=0\), the effective potential for \(\varphi\) becomes
\begin{align}
V(\varphi)
&=
8Kp^2\pi^2\cos^2\!\left(\frac{\beta\varphi}{2}\right),
\end{align}
and is extremized whenever
\begin{align}
V_{,\varphi}
&=
-4Kp^2\pi^2\beta\,\sin\!\big(\beta\varphi\big)
= 0
\quad\Longleftrightarrow\quad
\varphi_n = \frac{n\pi}{\beta},
\qquad
n \in \mathbb{N}_0.
\end{align}
In order to expand around a stable minimum of the potential we require
\begin{align}
V_{,\varphi\varphi}
&=
-4Kp^2\pi^2\beta^2\cos\!\big(\beta\varphi\big)
>0,
\end{align}
which implies that only odd values of \(n\) are allowed. For definiteness we take \(n=1\) from now on, so that
\begin{align}
\varphi_1 = \frac{\pi}{\beta},\quad
\Psi_0 = 0,
\end{align}
solves the equations of motion.

Expanding around this background as \(\varphi \to \varphi_1 + \chi\) and \(\Psi \to \Psi\), we obtain the following effective theory for chiral and Maxwell perturbations in \(1+1\) dimensions:
\begin{align}
\mathscr{L}\big[\chi,\Psi\big]
&=
-\frac{1}{2}\,\partial_\mu\chi\,\partial^\mu\chi
-\frac{1}{2}\,\partial_\mu\Psi\,\partial^\mu\Psi
-2e^2K\,\Psi^2\cos^2\!\left(\frac{\beta\chi}{2}\right)
-8Kp^2\pi^2\sin^2\!\left(\frac{\beta\chi}{2}\right),
\label{CM11}
\end{align}
whose equations of motion read
\begin{align}
\square\chi
+ e^2K\beta\,\Psi^2\sin\!\big(\beta\chi\big)
- 4Kp^2\pi^2\beta\,\sin\!\big(\beta\chi\big)
&= 0,\\
\square\Psi
- 4e^2K\,\cos^2\!\left(\frac{\beta\chi}{2}\right)\Psi
&= 0.
\end{align}
Close to \(\chi=0\) the reduced small--oscillation gaps of the two retained modes are
\begin{align}
m_\chi^2 = 4Kp^2\pi^2\beta^2 = \frac{2p^2}{L^2},\qquad m_\Psi^2 = 4e^2K.
\label{masses}
\end{align}
These scales should be read as gaps of the reduced \(1+1\) theory, not as particle masses of ordinary infinite-volume ChPT. No local \(U(1)\) gauge symmetry is broken by the reduction: the retained variable \(\Psi\) is a gauge-invariant finite-volume holonomy coordinate, and \(m_\Psi\) measures the curvature of the reduced energy along this gauge-invariant direction, rather than a Proca mass generated by a Higgs mechanism. Similarly, \(m_\chi\) arises from restricting the massless chiral theory to a sector with fixed transverse winding; it is a finite-volume energy cost within that topological sector, not an explicit chiral-symmetry-breaking pion mass. The full unreduced Chiral--Maxwell theory retains its gauge symmetry and should not be interpreted as ordinary infinite-volume ChPT with a massive photon.

The infinite-volume limit $L\to\infty$ is instructive. In that limit $m_\chi^2 = 2p^2/L^2 \to 0$: the chiral gap vanishes, consistently with the fact that in infinite-volume ChPT the pion is massless (no explicit symmetry breaking is present in our action). At the same time, although $m_\Psi^2 = 4e^2K$ formally remains finite, $\Psi$ itself ceases to be a distinct degree of freedom: Wilson-line holonomies are zero-modes of compact cycles and do not exist as normalizable modes in infinite volume. The entire reduced \(1+1\) description breaks down for $L\to\infty$, and with it any notion of a condensed holonomy branch. No Higgs mechanism or photon mass persists in the infinite-volume theory. This is the precise analogue of the Hosotani mechanism \cite{Hosotani83,AitkenChermanPoppitzYaffe2017}: on a compact circle the holonomy zero-mode acquires a nontrivial effective potential through quantum corrections, while in infinite volume the circle decompactifies and the holonomy degree of freedom disappears from the spectrum. Our $m_\Psi$ plays the role of the Hosotani-mechanism gap for the transverse holonomy, not a Proca mass for the photon.

We may now attempt to classically integrate out the chiral degree of freedom \(\chi\) by using its own equation of motion in a tree-level (saddle-point) approximation. Neglecting the dynamics of \(\chi\) and treating it as an auxiliary field turns its equation of motion into the constraint
\begin{align}
\bigl(e^2K\beta\,\Psi^2 - 4Kp^2\pi^2\beta\bigr)\,\sin\!\big(\beta\chi\big)
&= 0.
\end{align}
This constraint admits two branches. The first one is
\begin{align}
\sin\!\big(\beta\chi\big) &= 0
\quad\Longrightarrow\quad
\chi_n = \frac{n\pi}{\beta},
\qquad
n\in\mathbb{N}_0.
\end{align}
The second branch would instead impose
\begin{align}
e^2K\,\Psi^2 - 4Kp^2\pi^2 &= 0,
\end{align}
which corresponds to forcing the retained gauge coordinate to sit at a non--zero classical value. This is incompatible with our choice of \(\Psi\) as a fluctuation around a vanishing background, so we discard this possibility within the present small–oscillation framework.

We therefore focus on the first branch. To remain in the vicinity of the chosen stable background \(\varphi=\varphi_1\) we restrict to solutions with \(\varphi_1+\chi_n\) still at a minimum, and we may without loss of generality pick the trivial representative,
\begin{align}
\chi &= 0
\quad\Longleftrightarrow\quad
\varphi = \varphi_1.
\end{align}
With this choice, the effective equation for the reduced gauge fluctuation becomes a gapped Klein--Gordon equation,
\begin{align}
\square\Psi - m_\Psi^2\,\Psi &= 0,
\end{align}
with \(m_\Psi\) defined in \eqref{masses}. The saddle-point elimination of \(\chi\) thus produces a free gapped theory for the retained coordinate \(\Psi\) and no nontrivial self-interaction, which motivates a quantum treatment of the chiral sector in the next subsection.

\subsection{Coleman Duality and One-Loop Effective Action}\label{subsec:Coleman}

To obtain a genuinely self-interacting effective field theory for \(\Psi\), we reorganize the \(\chi\) sector in a fermionic language. In the purely bosonic \(1+1\) description, \(\chi\) is dimensionless, so the effective potential can contain an infinite tower of local operators without derivatives, with no obvious hierarchy under naive power counting. Coleman duality trades this tower for a fermionic model with only a small set of relevant and marginal operators, making the low-energy truncation for the reduced gauge-mode potential transparent.

We first rewrite the Lagrangian \eqref{CM11} as
\begin{align}
\mathscr{L}\big[\chi,\Psi\big]
&=
-\frac{1}{2}\,\partial_\mu\chi\,\partial^\mu\chi
+ \frac{m_\chi^2}{\beta^2}\left(\cos(\beta\chi)-1\right)
-\frac{1}{2}\,\partial_\mu\Psi\,\partial^\mu\Psi
- e^2K\,\Psi^2\cos(\beta\chi)
- e^2K\,\Psi^2,
\end{align}
with \(m_\chi^2\) defined in \eqref{masses}. In this form, the \(\chi\)–sector is exactly a sine–Gordon model with frequency \(\beta\).

Coleman duality \cite{Coleman1975,Mandelstam1975,Thirring1958} maps the sine–Gordon theory to the massive Thirring model through the standard dictionary
\begin{align}
-\frac{\beta}{2\pi}\,\epsilon^{\mu\nu}\partial_\nu\chi &\;\longleftrightarrow\; \bar\nu\gamma^\mu\nu,
&
\frac{m_\chi^2}{\beta^2}\cos(\beta\chi) &\;\longleftrightarrow\; -M\,\bar\nu\nu,
&
\frac{4\pi}{\beta^2}&=1+\frac{g}{\pi}.
\end{align}
We take $\epsilon^{01}=+1$ and the Dirac matrices satisfy $\{\gamma^\mu,\gamma^\nu\}=2\eta^{\mu\nu}$ with $\eta_{\mu\nu}=\mathrm{diag}(-1,+1)$.
Here \(\nu\) is a Dirac fermion, and
\begin{align}
  M \equiv \frac{8\,m_\chi}{\beta^2}
\end{align}
is the classical soliton mass, and \(g\) is the Thirring coupling. We use this dictionary as a low-energy matching prescription for the reduced EFT: finite renormalizations of \(M\) and of analytic local operators are absorbed into the definition of \(M\) and into the counterterms below. In our case the cosine appears multiplied by a \(\Psi\)–dependent prefactor, so in the dual description the fermion mass term becomes \(-M_{\text{eff}}(\Psi)\,\bar\nu\nu\) with
\begin{align}
M_{\text{eff}}(\Psi)
&=
M\left(1-\frac{e^2K\beta^2}{m_\chi^2}\,\Psi^2\right).
\label{Meff}
\end{align}
The theory can therefore be written as
\begin{align}
\mathscr{L}[\nu,\Psi]
&=
\bar{\nu}\big(i \gamma^\mu\partial_\mu  -M_{\text{eff}}(\Psi)\big)\nu
-\frac{g}{2}\big(\bar{\nu}\gamma^\mu\nu\big)\big(\bar{\nu} \gamma_\mu\nu\big)
-\frac{1}{2}\,\partial_\mu\Psi\,\partial^\mu\Psi
- e^2K\,\Psi^2.
\end{align}

To obtain the one-loop effective potential for \(\Psi\), we integrate out \(\nu\) as a free fermion with mass \(M_{\text{eff}}(\Psi)\). One may also make the Thirring coupling explicit by introducing an auxiliary field \(B_\mu\) via Hubbard--Stratonovich \cite{Stratonovich1957,Hubbard1959},
\begin{align}
-\frac{g}{2}\big(\bar{\nu}\gamma^\mu\nu\big)\big(\bar{\nu} \gamma_\mu\nu\big)
&\;\longrightarrow\;
B_\mu\big(\bar\nu\gamma^\mu\nu\big)
-\frac{1}{2g}B_\mu B^\mu,
\end{align}
so that
\begin{align}
\mathscr{L}[\nu, \Psi,B^\mu]
&=
\bar{\nu} \big(i\slashed\partial + \slashed{B} - M_\text{eff}(\Psi)\big)\nu
-\frac{1}{2}\,\partial_\mu\Psi\,\partial^\mu\Psi
- e^2K\,\Psi^2
-\frac{1}{2g}B_\mu B^\mu.
\end{align}
Integrating out \(\nu\) produces the determinant
\begin{align}
\Delta S[\Psi,B^\mu]
&=
-i \,\text{Tr} \log \big(i \slashed{\partial} + \slashed{B} - M_{\text{eff}}(\Psi)\big),
\end{align}
whose expansion in \(B_\mu\) begins as
\begin{align}
\Delta S[\Psi,B^\mu]
&=
-i \,\text{Tr} \log \big(i \slashed{\partial} - M_{\text{eff}}(\Psi)\big)
+\frac{i}{2}\,\text{Tr}\left(\frac{1}{i\slashed\partial-M_{\text{eff}}}\,\slashed B\right)^2
+\cdots.
\end{align}
In \(1+1\) dimensions the quadratic term induces
\begin{align}
\Delta\mathscr{L}[B^\mu]
&=
\frac{1}{2\pi}\,B_\mu B^\mu,
\end{align}
so the Gaussian integral over \(B_\mu\) is controlled by
\begin{align}
\mathscr{L}_{\text{eff}}[B_\mu]
&=
-\frac{1}{2} \left(\frac{1}{g} - \frac{1}{\pi}\right)B_\mu B^\mu.
\end{align}
In the regime of interest here (slowly varying, effectively homogeneous backgrounds for \(\Psi\) and a one-loop treatment of quantum corrections) this contributes only a \(\Psi\)-independent normalization. Equivalently, we are extracting the zero-momentum effective potential for the retained cavity gauge coordinate \(\Psi\); derivative corrections and any momentum-dependent dressing of the auxiliary sector are suppressed in this approximation and do not affect the stationary-point analysis below. Hence, within this approximation the entire \(\Psi\)-dependence resides in the standard fermionic determinant with mass \(M_{\text{eff}}(\Psi)\),
\begin{align}
i\,\text{Tr} \log\big(i \slashed{\partial} - M_{\text{eff}}(\Psi)\big)
&=
- \int d^2x \,\frac{M_{\text{eff}}^2(\Psi)}{4\pi}
\log \left( \frac{M_{\text{eff}}^2(\Psi)}{\mu^2} \right),
\end{align}
where \(\mu\) is the renormalization scale and scheme–dependent local terms have been absorbed into counterterms.

Putting everything together, the one–loop effective Lagrangian for \(\Psi\) takes the form
\begin{align}
\mathscr{L}_{\text{eff}}[\Psi]
&=
-\frac{1}{2}\,\partial_\mu\Psi\,\partial^\mu\Psi
- e^2K\,\Psi^2
+ \frac{1}{4\pi}\,M_\text{eff}^2(\Psi)\,
\log\!\left(\frac{M_\text{eff}^2(\Psi)}{\mu^2}\right)
- \delta m_\Psi^2(\mu)\,\Psi^2,
\end{align}
with \(M_{\text{eff}}(\Psi)\) given in \eqref{Meff} and \(\delta m_\Psi^2(\mu)\) a counterterm fixed by a renormalization condition.

We next specialize the one--loop effective theory to the trivial branch \(\Psi_{\mathrm{vac}} = 0\). We fix the renormalization scale at
\begin{align}
\mu &= M,
\end{align}
and impose the curvature condition
\begin{align}
V_{\mathrm{eff}}''(0) &= 4 e^2 K,
\end{align}
so that the curvature of the reduced gauge-mode potential at the origin matches the tree--level gap in \eqref{masses}. This fixes
\begin{align}
\delta m_\Psi^2(M)
&=
e^2 K\left(1 - \frac{M^2 \beta^2}{2\pi m_\chi^2}\right).
\end{align}
The form below corresponds to a definite finite subtraction prescription. Besides fixing the vacuum energy at \(\Psi=0\) and the curvature condition above, we choose the finite local quartic counterterm at the matching scale \(\mu=M\) to vanish. Equivalently, possible analytic \(\Psi^4\) contributions not fixed by the curvature condition are regarded as higher-order Wilson coefficients of the reduced EFT and are set to zero in the minimal scheme used to display the local branch structure. Varying this finite local coefficient deforms the quantitative location of the window but not the qualitative point that the logarithmic determinant can generate nontrivial local stationary branches.

Inserting this back into the effective Lagrangian yields the one--loop theory around the trivial branch,
\begin{align}
\mathscr{L}_{\mathrm{eff}}[\Psi]
&=
- \frac{1}{2}\,\partial_\mu \Psi\,\partial^\mu \Psi
- e^2 K\left(2 - \frac{M^2 \beta^2}{2\pi m_\chi^2}\right)\Psi^2
+ \frac{M^2}{4\pi}
\left(1 - \frac{e^2 K \beta^2}{m_\chi^2}\,\Psi^2\right)^{2}
\log\!\left[
\left(1 - \frac{e^2 K \beta^2}{m_\chi^2}\,\Psi^2\right)^{2}
\right].
\label{eq:LeffPsi_1loop}
\end{align}

It is convenient to use the dimensionless combination
\begin{align}
\tau \equiv \frac{e^2 K \beta^2}{m_\chi^2}\,\Psi^2
= \frac{e^2}{4\pi^2 p^2}\,\Psi^2 \ge 0,
\label{eq:t_def}
\end{align}
which measures the proximity to the point where the effective fermion mass ${M_{\mathrm{eff}}(\Psi)=M(1-\tau)}$ in \eqref{Meff} would vanish, where in the second equality we used $m_\chi^2=4Kp^2\pi^2\beta^2$ from \eqref{masses}.
Since the logarithm originates from integrating out a fermion of mass \(M_{\mathrm{eff}}(\Psi)\), the single-field description is reliable only when the integrated fermion remains parametrically heavy compared to the retained scales and one stays away from the vicinity where \(M_{\mathrm{eff}}(\Psi)\) approaches zero, i.e.\ away from $\tau\simeq 1$, and when the logarithm is not parametrically large. Consequently the effective potential below is used to diagnose the local stationary structure and the onset of a reduced-mode instability; it is not a controlled extrapolation to arbitrarily large values of \(\Psi\).

\begin{figure}[t]
\centering
\includegraphics[width=\linewidth]{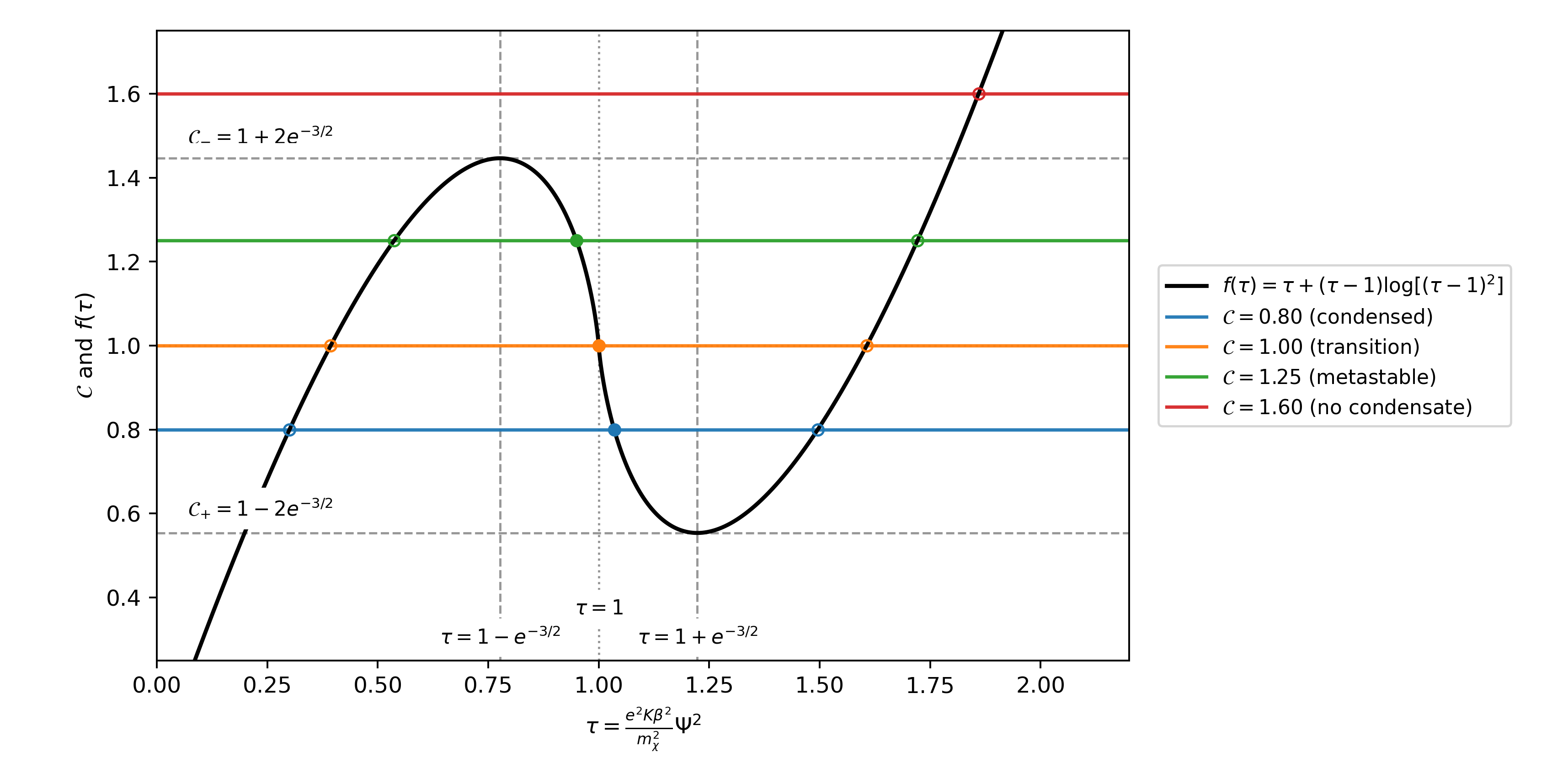}
\caption{\scriptsize Gap-function representation of the stationary-point condition \(\mathcal{C}=f(\tau_\star)\), with \(f(\tau)=\tau+(\tau-1)\log[(\tau-1)^2]\). The vertical markers indicate \(\tau=1\) and \(\tau_\pm=1\pm e^{-3/2}\). The dashed horizontal levels indicate \(\mathcal{C}_\pm=1\mp 2e^{-3/2}\) and the intermediate value \(\mathcal{C}=1\). Colored horizontal lines show representative values of \(\mathcal{C}\); their intersections with \(f(\tau)\) give the candidate solutions \(\tau_\star\). Filled markers correspond to intersections within the stability band \(\tau_-<\tau_\star<\tau_+\), while open markers lie outside it.}
\label{fig:gap}
\end{figure}

We also consider the possibility of a locally condensed branch in which the effective potential for the reduced gauge coordinate \(\Psi\) develops nontrivial stationary minima \(\Psi_\star\neq 0\). In terms of the dimensionless variable $\tau$ in \eqref{eq:t_def}, such extrema correspond to stationary points $\tau_\star>0$. We then introduce the dimensionless control parameter
\begin{align}
\mathcal{C} &\equiv \frac{4\pi m_\chi^2}{\beta^2 M^2}
= \frac{\pi\beta^2}{16}.
\end{align}
The effective potential can be rewritten as
\begin{align}
V_{\mathrm{eff}}(\tau)
&=
\frac{M^2}{4\pi}
\Big[
2(\mathcal{C}-1)\,\tau
- (1-\tau)^2 \log\!\big((1-\tau)^2\big)
\Big],
\end{align}
with \(V_{\mathrm{eff}}(0)=0\) by construction. Extrema with \(\Psi_\star\neq 0\) correspond to solutions with \(\tau_\star>0\) of the gap equation
\begin{align}
\frac{dV_{\mathrm{eff}}}{d\tau}\bigg|_{\tau=\tau_\star}&=0,
\end{align}
which yields
\begin{align}
\mathcal{C}
&=
\tau_\star
+ (\tau_\star-1)\,\log\!\big[(\tau_\star-1)^2\big].
\label{eq:gap-photon}
\end{align}
Defining \(f(\tau)\equiv \tau+(\tau-1)\log[(\tau-1)^2]\), the stationarity condition in \eqref{eq:gap-photon} can be written as
\begin{align}
\mathcal{C}=f(\tau_\star).
\end{align}
This formulation is useful because the stationary-point structure reduces to a geometric statement about intersections between the graph of \(f(\tau)\) and horizontal lines \(\mathcal{C}=\mathrm{const}\). The function \(f(\tau)\) has two critical points at
\begin{align}
\tau_\pm = 1\pm e^{-3/2},
\end{align}
which map to the extrema
\begin{align}
\mathcal{C}_\pm = f(\tau_\pm)= 1\mp 2e^{-3/2},\qquad (\mathcal{C}_+,\mathcal{C}_-)\simeq (0.554,\,1.446).
\end{align}
As a consequence, for \(\mathcal{C}<\mathcal{C}_+\) or \(\mathcal{C}>\mathcal{C}_-\) the equation \(\mathcal{C}=f(\tau_\star)\) admits a single real solution, whereas for \(\mathcal{C}_+<\mathcal{C}<\mathcal{C}_-\) it admits three real solutions. Moreover, the curvature criterion derived below selects the locally stable branch within the band \(\tau_-\!<\tau_\star<\tau_+\). Figure~\ref{fig:gap} summarizes this structure and makes the multiplicity of solutions and the stability band manifest.

Figure~\ref{fig:gap} should be read as a compact stationary-point diagram for $V_{\mathrm{eff}}(\Psi)$. For a fixed value of the microscopic ratio $\mathcal{C}$, the intersections of the horizontal line $\mathcal{C}=\mathrm{const}$ with the curve $f(\tau)$ give all candidate extrema $\tau_\star$ and hence $\Psi_\star^2\propto \tau_\star$. The vertical markers at $\tau=1$ and $\tau=\tau_\pm$ isolate the region where the fermionic determinant is close to its would-be zero, $M_{\mathrm{eff}}\to0$, and where the one-mode treatment becomes most delicate. Within the stable band $\tau_-<\tau_\star<\tau_+$ (filled markers) the extremum is a genuine local minimum of the one-loop effective potential for the reduced coordinate; outside it (open markers) it is a maximum or an unstable saddle. This is a local statement inside the domain where the single-field EFT is meaningful. To decide whether the locally stable stationary point is favored relative to the trivial branch in this domain, we next compare its energy to the trivial branch by evaluating a dimensionless version of the potential.

It is also useful to note that \eqref{eq:gap-photon} admits a closed-form solution. Writing \(\Delta \equiv \mathcal{C}-1\) and \(y\equiv \tau-1\), the equation becomes
\begin{align}
\Delta = y\left(1+2\log|y|\right).
\end{align}
For the branch \(y>0\) (extrema to the right of \(\tau=1\)), define the auxiliary variable \(v\equiv 1+2\log y\), which brings the equation to
\begin{align}
\frac{v}{2}\,e^{v/2} = \frac{\Delta e^{1/2}}{2},
\end{align}
so that, in terms of the Lambert function \(W_k\)\footnote{The Lambert \(W\) function is defined as the (multi)inverse of \(w\mapsto w e^{w}\): by definition \(w=W(z)\) iff \(w\,e^{w}=z\). It has countably many branches \(W_k\) (\(k\in\mathbb{Z}\)); on the real axis the principal branch \(W_0\) is real for \(z\ge -e^{-1}\), and a second real branch \(W_{-1}\) exists on \(-e^{-1}\le z<0\). A useful identity is \(W'(z)=\dfrac{W(z)}{z\,(1+W(z))}\) (for \(z\neq 0\) and \(W(z)\neq -1\)), so equations of the form \(x e^{x}=a\) are solved by \(x=W(a)\). See \cite{CorlessLambertW,NISTHandbook}.},
\begin{align}
y = \frac{\Delta}{2\,W_k\!\left(\dfrac{\Delta e^{1/2}}{2}\right)} \Longrightarrow \tau = 1 + \frac{\mathcal{C}-1}{2\,W_k\!\left(\dfrac{(\mathcal{C}-1)e^{1/2}}{2}\right)}.
\end{align}
Different real branches $k$ (typically $W_0$ and $W_{-1}$) reproduce the different extrema of the effective potential for a fixed $\mathcal{C}$; here $k$ is the branch label of $W$ and should not be confused with the cavity wavenumber used elsewhere. An analogous construction holds for $y<0$ (solutions to the left of $\tau=1$).

The real branches \(W_0\) and \(W_{-1}\) coalesce at the branch point \(z=-1/e\), where \(W_0(-1/e)=W_{-1}(-1/e)=-1\). Since in the present case \(z=(\mathcal{C}-1)e^{1/2}/2\), this branch point corresponds to \(\mathcal{C}=\mathcal{C}_+\). On the \(\tau>1\) side, two real solutions exist only for \(\mathcal{C}_+<\mathcal{C}<1\); at \(\mathcal{C}=1\) one solution approaches the limiting point \(\tau\to1^+\), while for \(\mathcal{C}>1\) only one solution remains on this side. The full stationary-point count quoted above is recovered only after including also the \(\tau<1\) branch.

The stability of a given extremum is controlled by the curvature with respect to \(\Psi\). Using the gap equation \eqref{eq:gap-photon} to simplify \(V_{\mathrm{eff}}''\) at a stationary point gives
\begin{align}
V_{\mathrm{eff}}''(\Psi_\star)
&=
-\,\frac{2 e^2 K\,M^2\beta^2}{\pi m_\chi^2}\;
\tau_\star\;\Big(\log[(\tau_\star-1)^2] + 3\Big).
\end{align}
Hence a non–trivial extremum is a local minimum if and only if
\begin{align}
|\tau_\star-1| < e^{-3/2},
\end{align}
i.e.\ in the band
\begin{align}
1-e^{-3/2}\simeq 0.777
<
\tau_\star
<
1+e^{-3/2}\simeq 1.223.
\end{align}
At the band edges \(\tau=\tau_\pm\) one has \(\mathcal{C}=\mathcal{C}_\pm\) as above.
For \(\mathcal{C}_+<\mathcal{C}<\mathcal{C}_-\) the curve \(\mathcal{C}=f(\tau)\) admits three positive solutions; only the intermediate one lies inside the stability band and corresponds to a genuine nontrivial minimum.

\begin{figure}[t]
\centering
\includegraphics[width=\linewidth]{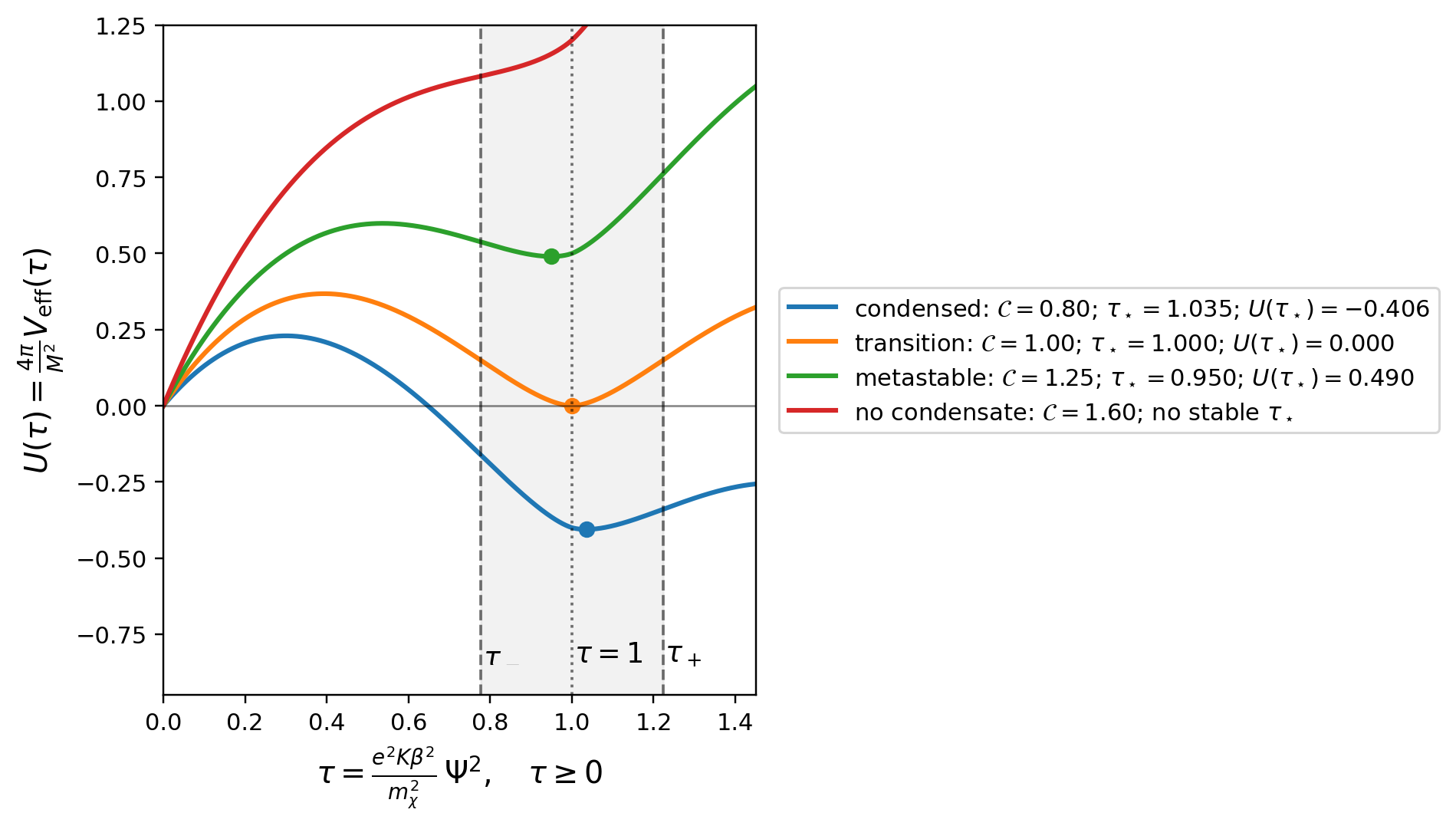}
\caption{\scriptsize Dimensionless effective potential \(U(\tau)=(4\pi/M^2)V_{\mathrm{eff}}(\tau)\) plotted versus \(\tau=(e^2K\beta^2/m_\chi^2)\Psi^2\ge 0\), for representative values of \(\mathcal{C}\) in the nontrivial local-branch, transition, metastable, and trivial-branch regimes. The horizontal baseline is \(U=0\). The vertical markers indicate \(\tau=1\) (dotted) and \(\tau=1\mp e^{-3/2}\) (dashed), as discussed in the text. Filled points mark the locally stable stationary point \(\tau_\star\) (when it exists) obtained from the gap equation and the stability criterion; the legend reports the corresponding values of \(\tau_\star\) and \(U(\tau_\star)\). The one-loop EFT is controlled in the region where \(M_{\mathrm{eff}}(\Psi)=M|1-\tau|\gg m_\Psi\), i.e.\ when \(\tau\) is bounded away from \(\tau=1\) by a margin \(\gg m_\Psi/M\) (see Sec.~\ref{subsec:validity}). The locally stable stationary points lie in \(\tau_\star\in(\tau_-,\tau_+)\) and the one-loop analysis is self-consistent there provided \(m_\Psi\ll M\). The apparent descent of the curves for \(\tau\gg 1\) has no physical significance for the full theory: it is an artefact of extrapolating the EFT beyond its domain of validity, not a global instability. The figure is used to characterize the local branch structure within the controlled regime.}
\label{fig:Ut}
\end{figure}

To compare the nontrivial local minimum with the trivial branch, we use the dimensionless potential
\begin{align}
U(\tau)&\equiv \frac{4\pi}{M^2}V_{\mathrm{eff}}(\tau)
=
2(\mathcal{C}-1)\,\tau
- (1-\tau)^2 \log\!\big[(1-\tau)^2\big],
\end{align}
with \(U(0)=0\). One finds that for \(\mathcal{C}_+ \lesssim \mathcal{C}<1\) the locally stable solution in the band can satisfy \(U(\tau_\star)<0\), yielding two degenerate reduced-mode minima at \(\Psi=\pm \Psi_\star\). At \(\mathcal{C}=1\) the stable nontrivial extremum approaches \(\tau_\star\to 1\) and becomes degenerate with the trivial branch, while for \({1<\mathcal{C}\lesssim \mathcal{C}_-}\) it persists but is higher in energy than \(\Psi=0\) (metastable). This branch structure is summarized in Figure~\ref{fig:Ut}.

Figure~\ref{fig:Ut} adds the missing energetic information inside the reduced description. The plotted quantity \(U(\tau)\) is normalized so that \(U(0)=0\) corresponds to the trivial branch \(\Psi=0\). Within the local domain where the one-loop reduced EFT is under control, a locally stable stationary point with \(U(\tau_\star)<0\) indicates that the reduced gauge coordinate develops a pair of energetically favored local branches at \(\Psi=\pm\Psi_\star\). For $\mathcal{C}_+<\mathcal{C}<1$, the locally stable nontrivial stationary point lies below the $\Psi=0$ baseline and is therefore energetically favored within the reduced EFT. As $\mathcal{C}$ is increased toward $\mathcal{C}=1$ from below, this stationary point approaches $\tau_\star\to1$ and becomes degenerate with the trivial branch. For $1<\mathcal{C}<\mathcal{C}_-$, the same local stationary point persists but lies above the $\Psi=0$ branch and is therefore metastable. Near $\mathcal{C}_-$ it merges with an unstable branch and disappears. The vertical markers at $\tau=1$ and $\tau=1\mp e^{-3/2}$ indicate where the determinant mass $M_{\mathrm{eff}}=M(1-\tau)$ is suppressed and the loop potential is most sensitive to additional modes.

\newpage
\vspace*{-1cm}

\subsection{Domain of Validity and Large-Field Behavior}
\label{subsec:validity}

The one-loop effective potential \eqref{eq:LeffPsi_1loop} was derived by integrating out a fermion of mass $M_{\mathrm{eff}}(\Psi) = M(1-\tau)$. The calculation is reliable only when this fermion remains parametrically heavy compared to all retained energy scales, i.e.\ when
\begin{align}
|M_{\mathrm{eff}}(\Psi)| = M\,|1-\tau| \gg m_\Psi,
\label{eq:validity_condition}
\end{align}
or equivalently when $\tau$ is bounded away from $\tau = 1$ on a scale set by $m_\Psi/M$. This is precisely the regime in which the single-field EFT description is self-consistent.

In practice, the one-loop determinant depends on $M_{\mathrm{eff}}^2 = M^2(1-\tau)^2$, so the calculation is mathematically well-defined for all $\tau \neq 1$. What breaks down is the physical validity of the expansion: as $\tau\to 1$ from either side, $|M_{\mathrm{eff}}|=M|1-\tau|$ approaches zero, making the chiral sector light rather than heavy. Once $|M_{\mathrm{eff}}|$ becomes comparable to $m_\Psi$, the hierarchy used to justify integrating out the chiral sector collapses, and higher-loop corrections, operator mixing, and modes not retained in the single-field truncation all become equally important. It is this vanishing of $|M_{\mathrm{eff}}|$ near $\tau=1$, not the sign of $M_{\mathrm{eff}}$, that marks the boundary of reliability. Consequently, the apparent descent of $V_{\mathrm{eff}}(\tau)$ for $|\tau-1|\gg e^{-3/2}$ visible in Figure~\ref{fig:Ut} is \emph{not} a physical instability of the full Chiral--Maxwell theory; it is an artefact of extrapolating the one-loop determinant into the regime where it no longer approximates the exact effective potential.

This situation is familiar from other one-loop effective potentials. In the Coleman--Weinberg mechanism for a massless scalar \cite{Coleman1973}, the one-loop potential is only meaningful in the weak-field regime where higher-loop contributions are suppressed; its apparent large-field behavior has no physical significance once loop counting breaks down. More directly, the massive Thirring / sine--Gordon correspondence itself is valid for a specific range of coupling $\beta^2$ \cite{Coleman1975}, and the fermionic determinant computed here inherits the same range of applicability. Within this range, the local stationary structure and the local-condensation window derived above are well-defined and unambiguous conclusions of the one-loop analysis.

The quantitative criterion is: the effective fermion mass at the stationary point,
\begin{align}
M_{\mathrm{eff}}(\tau_\star) = M\,|\tau_\star - 1|,
\label{eq:Meff_at_minimum}
\end{align}
must satisfy $M|\tau_\star - 1| \gg m_\Psi$ for the calculation to be controlled. The locally stable nontrivial minimum lies in $\tau_\star \in (\tau_-, \tau_+)$, so $|\tau_\star - 1| \in (0, e^{-3/2})$, which gives $M_{\mathrm{eff}}(\tau_\star) \in (0, Me^{-3/2})$. The one-loop analysis is self-consistent at the minimum provided
\begin{align}
\frac{m_\Psi}{M} \ll \tau_\star - 1 \qquad (\tau_\star > 1)
\quad \text{or} \quad
\frac{m_\Psi}{M} \ll 1 - \tau_\star \qquad (\tau_\star < 1),
\label{eq:control_condition}
\end{align}
which is realised in the natural hierarchy $m_\Psi \ll M$. For the most energetically preferred minima (close to $\tau_+ \approx 1.22$ or $\tau_- \approx 0.78$), $|\tau_\star - 1| \sim e^{-3/2} \approx 0.22$, so the requirement $m_\Psi/M \ll 0.22$ is comfortably satisfied whenever $m_\Psi \lesssim M/10$. Within this regime, both the local minimum and the energetic comparison $U(\tau_\star) < 0$ are physically meaningful one-loop results.

The same logarithmic structure also makes the limitations of the one-mode description explicit. The nontrivial local minimum lies close to \(\tau_\star\simeq 1\), where \(M_{\mathrm{eff}}(\Psi_\star)=M(1-\tau_\star)\) is suppressed and the separation of scales used to integrate out the chiral sector becomes marginal. In that regime the potential should be regarded as a Landau--Ginzburg indicator of an incipient finite-volume instability of the reduced gauge mode, not as a proof of an absolutely stable global vacuum of the full theory. A quantitatively controlled analysis near the critical point requires keeping both degrees of freedom dynamical.

It is also useful to contrast this with the original two-field theory. The classical potential in the \((\varphi,\Psi)\) language,
\begin{align}
V_{\text{cl}}(\varphi,\Psi)
=
2e^2K\,\Psi^2\sin^2\!\Big(\frac{\beta\varphi}{2}\Big)
+ 8Kp^2\pi^2\cos^2\!\Big(\frac{\beta\varphi}{2}\Big),
\end{align}
is minimized by \(\Psi=0\) and \(\varphi=(2n+1)\pi/\beta\), with no competing homogeneous minimum at \(\Psi\neq 0\). The appearance of nontrivial minima in \(V_{\mathrm{eff}}(\Psi)\) is therefore a genuinely quantum effect driven by the chiral sector encoded through the fermionic determinant. Within the scope of the present work we use this one-mode potential to identify the parameter window where such a local instability is expected, while a full coupled analysis of the two-field dynamics is left for future work.

Having derived an effective description for the lowest reduced gauge mode by integrating out the chiral sector, we now consider the complementary limit. Namely, we treat the gauge-induced scalar $\Psi$ as the heavy degree of freedom and integrate it out, obtaining a one-loop corrected sine--Gordon theory for the chiral mode $\chi$.

\section{One-Loop Chiral Sine--Gordon EFT}\label{Sec:EFTChi}

In the complementary hierarchy---where the reduced gauge mode $\Psi$ is heavy around a chosen minimum---we integrate it out at one loop and obtain a sine--Gordon-type EFT for the chiral mode $\chi$ with an explicit nonanalytic deformation of the potential. The sine--Gordon structure here should not be confused with the usual massive-pion sine--Gordon reduction produced by an explicit chiral symmetry breaking term. In the present construction it descends from the kinetic energy of the massless chiral field evaluated on the equivariant sector with rigid transverse winding.

Equivalently, the periodic potential for the reduced coordinate is not inserted as a microscopic chiral-symmetry-breaking operator. It is the finite-volume energy functional obtained after fixing the equivariant transverse winding data \(F=p\mathfrak z\) and \(G=p\mathfrak y\) and allowing only the longitudinal profile to remain dynamical. Thus the ``sine--Gordon mass'' is a gap within this restricted topological sector, not an explicit pion mass in the underlying ChPT Lagrangian.

The mechanism is kinematic, not dynamic. It is the direct analogue of Kaluza--Klein mass generation: a massless field in $d+1$ dimensions, when compactified on a circle of radius $L$ with winding number $p$, acquires a Kaluza--Klein gap $\sim p/L$ from the quantized momentum in the compact direction. Here the transverse winding supplies the analogous quantized momentum to the longitudinal profile $H(t,x)$, resulting in $m_\chi^2 = 2p^2/L^2$ from \eqref{masses}. No mass operator is added to the action, and the underlying $SU(2)$ chiral symmetry $U \to V U$ (for constant $V \in SU(2)$) is not broken by this procedure: under such a transformation the reduced profile $H(t,x)$ shifts, but the effective potential $V(\varphi)$ possesses the discrete shift symmetry $\varphi \to \varphi + 2\pi/\beta$ inherited from the gauge-invariant winding structure. No Goldstone mode is eaten; the gap is a topological sector cost rather than a consequence of spontaneous or explicit symmetry breaking.

As already made explicit when rewriting \eqref{CM11}, the $\chi$ sector is naturally normalized as a sine--Gordon theory with frequency $\beta$, and the relevant mass scale is \(m_\chi^2\) in \eqref{masses}. The key ingredient in this limit is the $\chi$--dependent mass of the $\Psi$ mode,
\begin{align}
m_\Psi^{2}(\chi)
&= 4e^{2}K\,\cos^{2}\!\left(\frac{\beta\chi}{2}\right)
 = 2e^{2}K\big(1+\cos\beta\chi\big).\label{eq:mPsi_of_chi}
\end{align}
It will be useful to parameterize the strength of the loop-induced deformation by the dimensionless ratio
\begin{align}
\alpha \equiv \frac{e^2K\beta^2}{4\pi m_\chi^2}
= \frac{e^2}{16\pi^3 p^2},
\label{eq:alpha_def}
\end{align}
where in the second equality we used \eqref{masses}. The single-field EFT is reliable in the regime $\alpha<1$.
When $m_\Psi(\chi)$ is large compared to the characteristic scales of the $\chi$ fluctuations, we may integrate out $\Psi$ at Gaussian level and organize the result in a derivative expansion in $\chi$. At leading order in this expansion, the one-loop contribution depends only on $m_\Psi^2(\chi)$. The bosonic determinant in \(1+1\) dimensions gives
\begin{align}
\Delta V(\chi)
&= \frac{1}{2}\int\!\frac{d^2k}{(2\pi)^2}
  \log\big(k^2 + m_\Psi^2(\chi)\big)
 = \frac{1}{8\pi}\,m_\Psi^2(\chi)\,\log\frac{m_\Psi^2(\chi)}{\mu^2}
  + \text{(polynomial counterterms)},
\end{align}
where $\mu$ is a renormalization scale and the nonlogarithmic pieces are absorbed into a renormalization of tree-level parameters.

To remove scheme-dependent constants and avoid carrying an explicit $\mu$, we adopt a threshold-matching prescription. We choose as reference one of the stable minima at
\begin{align}
\chi_\star = \frac{2\pi n}{\beta},\qquad n\in\mathbb Z,
\label{eq:chi_star_min}
\end{align}
where $\cos(\beta\chi_\star/2)=1$ and therefore
\begin{align}
m_\Psi^2(\chi_\star) = 4e^2K.
\label{eq:mPsi_at_chistar}
\end{align}
Fixing the finite part of the counterterms such that the vacuum energy at $\chi_\star$ vanishes, and referencing the logarithm to the physical threshold \eqref{eq:mPsi_at_chistar}, the effective Lagrangian for the light mode takes the form
\begin{align}
\mathscr L_{\rm eff}[\chi]
&= -\frac{1}{2}\partial_\mu\chi\,\partial^\mu\chi - V_{\rm eff}(\chi),
\end{align}
with
\begin{align}
V_{\rm eff}(\chi)
&= \frac{m_\chi^2}{\beta^2}\big(1-\cos\beta\chi\big)
 + \frac{1}{8\pi}\,m_\Psi^2(\chi)\,
   \log\!\frac{m_\Psi^2(\chi)}{m_\Psi^2(\chi_\star)}.
\end{align}
Using \eqref{eq:mPsi_of_chi} and \eqref{eq:mPsi_at_chistar}, this becomes
\begin{align}
V_{\rm eff}(\chi)
&= \frac{m_\chi^2}{\beta^2}\big(1-\cos\beta\chi\big)
 + \frac{e^2K}{2\pi}\,\cos^2\!\left(\frac{\beta\chi}{2}\right)\,
   \log\!\left[\cos^2\!\left(\frac{\beta\chi}{2}\right)\right].
\label{eq:Veff_chi_explicit}
\end{align}

The stationary condition follows from $V'_{\rm eff}(\chi_\star)=0$. Differentiating \eqref{eq:Veff_chi_explicit} gives
\begin{align}
V'_{\rm eff}(\chi)
&=
\sin(\beta\chi)\left[
\frac{m_\chi^2}{\beta}
- \frac{e^2 K\beta}{4\pi}
\left(
\log\!\left(\cos^2 \frac{\beta\chi}{2}\right) + 1
\right)
\right].
\end{align}
There are two branches, namely
\begin{align}
\sin(\beta\chi_\star)&=0
\end{align}
and
\begin{align}
\frac{m_\chi^2}{\beta}
&= \frac{e^2 K\beta}{4\pi}
\left(
\log\!\left(\cos^2 \frac{\beta\chi_\star}{2}\right) + 1
\right).
\label{eq:branch_bracket0}
\end{align}
The first branch selects the discrete set
\begin{align}
\chi_\star = \frac{n\pi}{\beta},\qquad n\in\mathbb Z.
\label{eq:chis_st_extrema}
\end{align}
The second branch corresponds to additional extrema with $\sin(\beta\chi_\star)\neq 0$. Solving \eqref{eq:branch_bracket0} for $\cos^2(\beta\chi_\star/2)$ yields
\begin{align}
\cos^2\frac{\beta\chi_\star}{2}
&= \exp\!\left(\frac{4\pi m_\chi^2}{e^2K\beta^2}-1\right),
\label{eq:cos2_solution}
\end{align}
which is compatible with $0\le\cos^2\le 1$ only if
\begin{align}
m_\chi^2 \le \frac{e^2K\beta^2}{4\pi}.
\label{eq:cond_additional_extrema}
\end{align}

Here we restrict to the complementary regime
\begin{align}
m_\chi^2 > \frac{e^2K\beta^2}{4\pi},
\label{eq:cond_stable_minima}
\end{align}
for which \eqref{eq:branch_bracket0} has no solution (since the right-hand side of \eqref{eq:cos2_solution} would exceed one), so the only extrema are those in \eqref{eq:chis_st_extrema}. This choice also matches the heavy-mode logic: when \eqref{eq:cond_additional_extrema} holds, the extra extrema occur where $\cos^2(\beta\chi/2)$ is exponentially small and hence \(m_\Psi^2(\chi)\) in \eqref{eq:mPsi_of_chi} is suppressed, invalidating the assumption that \(\Psi\) is heavy.

The second derivative can be written as
\begin{align}
V_{\rm eff}''(\chi)
&=
\cos(\beta\chi)
\left[
m_\chi^2 - \frac{e^2 K \beta^2}{4\pi}
\left(
\log\!\left(\cos^2 \frac{\beta\chi}{2}\right) + 2
\right)
\right]
+ \frac{e^2 K \beta^2}{4\pi}.
\label{eq:Vsecond_general}
\end{align}
At the even points
\begin{align}
\chi_\star = \frac{2\pi n}{\beta},
\qquad n\in\mathbb Z,
\label{eq:even_points}
\end{align}
one has $\cos(\beta\chi_\star)=1$ and $\cos^2(\beta\chi_\star/2)=1$, so the logarithm vanishes and the small-oscillation gap is
\begin{align}
V_{\rm eff}''(\chi_\star)=m_\chi^2-\frac{e^2K\beta^2}{4\pi}.
\label{eq:Mchi2}
\end{align}
These are true minima provided
\begin{align}
V_{\rm eff}''(\chi_\star) > 0
\quad\Longleftrightarrow\quad
m_\chi^2 > \frac{e^2K\beta^2}{4\pi},
\label{eq:stability_condition}
\end{align}
which is precisely \eqref{eq:cond_stable_minima}.

At the odd points
\begin{align}
\chi_\star = \frac{(2n+1)\pi}{\beta},
\label{eq:odd_points}
\end{align}
one has $\cos(\beta\chi_\star)=-1$ and $\cos(\beta\chi_\star/2)=0$. While the combination
$\cos^2(\beta\chi/2)\log[\cos^2(\beta\chi/2)]$ tends to zero as $\chi\to\chi_\star$, the curvature \eqref{eq:Vsecond_general} diverges to $-\infty$. These points are therefore sharp maxima. Moreover, \(m_\Psi^2(\chi)\to 0\) there by \eqref{eq:mPsi_of_chi}, so the \(\Psi\) mode ceases to be heavy and the single-field EFT cannot be quantitatively reliable in their vicinity.

\begin{figure}[t]
\centering
\includegraphics[width=\linewidth]{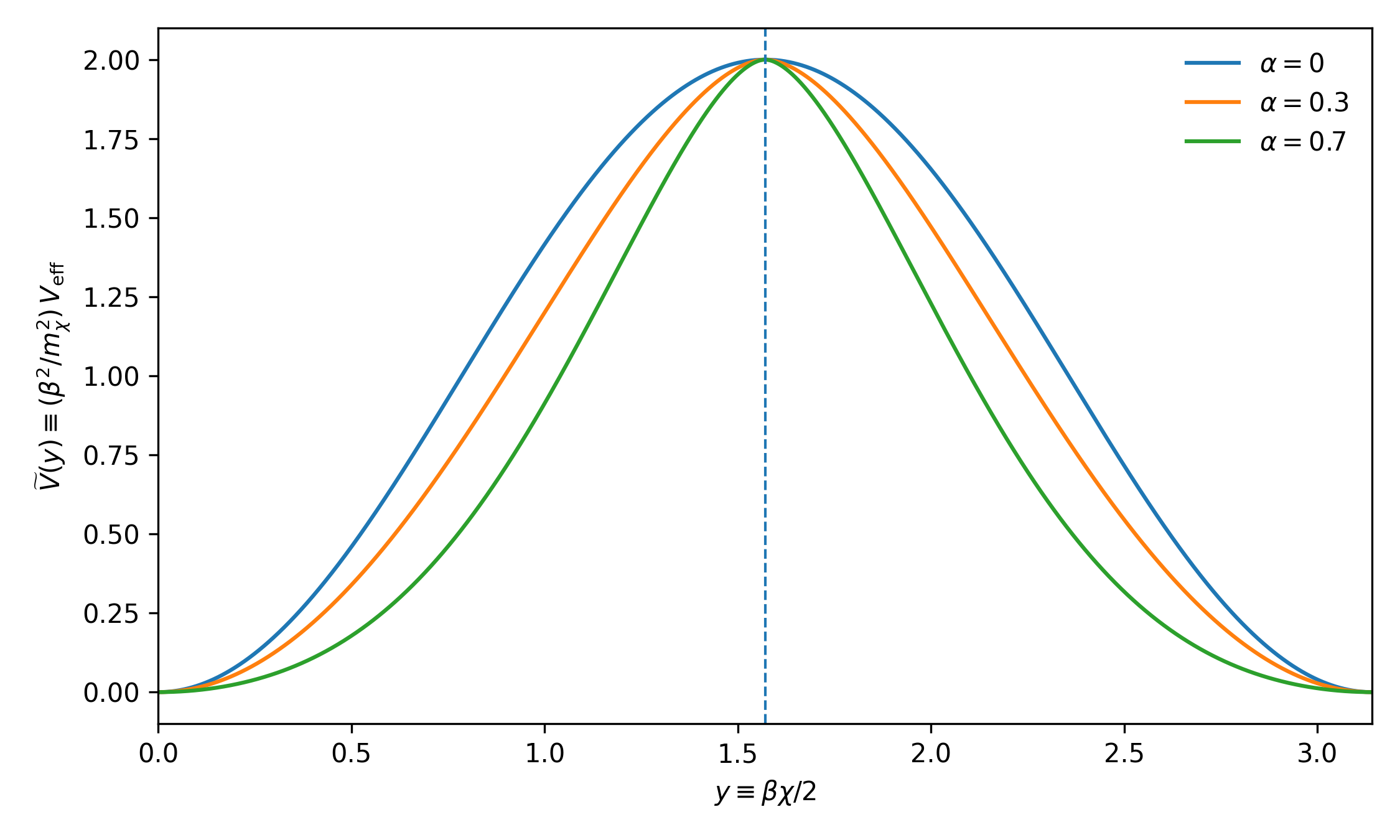}
\caption{\scriptsize One-loop deformation of the sine--Gordon landscape. We plot the dimensionless effective potential \(\widetilde V(y)\equiv (\beta^2/m_\chi^2)\,V_{\rm eff}(\chi)\) associated with \eqref{eq:Leff_chi_final} as a function of \(y\equiv\beta\chi/2\), for representative values of \(\alpha\equiv e^2K\beta^2/(4\pi m_\chi^2)\) in the stable regime \eqref{eq:stability_condition}. The vacua remain at \(y=n\pi\) (i.e.\ \(\chi=2\pi n/\beta\)), while the dashed vertical marker highlights the odd points \(y=(2n+1)\pi/2\) where \(m_\Psi^2(\chi)\to 0\) and the single-field description becomes marginal.}
\label{fig:Veffchi}
\end{figure}

Collecting the above results, after integrating out the heavy mode $\Psi$ at Gaussian level and performing threshold matching at \eqref{eq:chi_star_min}, the light degree of freedom $\chi$ is governed by the one-loop-corrected sine--Gordon Lagrangian
\begin{align}
\mathscr L_{\rm eff}[\chi]
&= -\frac{1}{2}\partial_\mu\chi\,\partial^\mu\chi
   -\frac{m_\chi^2}{\beta^2}\big(1-\cos\beta\chi\big)
   -\frac{e^2K}{2\pi}\,\cos^2\!\left(\frac{\beta\chi}{2}\right)\,
    \log\!\left[\cos^2\!\left(\frac{\beta\chi}{2}\right)\right].
\label{eq:Leff_chi_final}
\end{align}
To visualize the loop-induced deformation of the chiral landscape encoded in \eqref{eq:Leff_chi_final}, we define the dimensionless field variable \(y\equiv \beta\chi/2\) and the dimensionless potential \(\widetilde V(y)\equiv (\beta^2/m_\chi^2)\,V_{\rm eff}(\chi)\). Figure~\ref{fig:Veffchi} displays \(\widetilde V(y)\) for representative values of the control ratio \(\alpha\equiv e^2K\beta^2/(4\pi m_\chi^2)\) within the stable regime \eqref{eq:stability_condition}. The special points \(y=(2n+1)\pi/2\) correspond to \eqref{eq:odd_points}, where \(m_\Psi^2(\chi)\to 0\) and the hierarchy underlying the single-field EFT becomes marginal.

Figure~\ref{fig:Veffchi} illustrates two universal features of this threshold-matched one-loop EFT. First, the location of the vacua is unchanged by construction: the minima remain at $y=n\pi$ so that the topological sectors and the kink winding are preserved. Second, the loop term introduces a nonpolynomial deformation of the barrier: it lowers the potential away from the midpoint and produces a sharp nonanalytic curvature near the points where the integrated gauge mode becomes light. At the same time, it lowers the curvature at the stable vacua according to \eqref{eq:Mchi2}. The dashed marker at $y=(2n+1)\pi/2$ highlights the points where $m_\Psi^2(\chi)\to0$ and the heavy-mode assumption used to integrate out $\Psi$ fails; the plot should therefore be interpreted as reliable in neighborhoods of $y=n\pi$ and as qualitative near the dashed lines.

This EFT is reliable for fluctuations around the stable vacua \eqref{eq:even_points} as long as \(m_\Psi^2(\chi)\) in \eqref{eq:mPsi_of_chi} remains parametrically large compared to the light gap \eqref{eq:Mchi2}.

Next we turn to the topological sector of the chiral EFT and look for kink configurations. The effective potential \eqref{eq:Veff_chi_explicit} is $2\pi/\beta$ periodic and, when \eqref{eq:cond_stable_minima} holds, it admits a discrete set of stable vacua at
\begin{align}
\chi_{\rm vac}^{(n)}=\frac{2\pi n}{\beta},\qquad n\in\mathbb{Z},
\end{align}
with $V_{\rm eff}\!\left(\chi_{\rm vac}^{(n)}\right)=0$ by construction of the threshold matching. Between neighboring vacua the potential develops a sharp barrier near the odd points \eqref{eq:odd_points}, where $m_\Psi^2(\chi)\to 0$ and the hierarchy of scales underlying the single-field EFT becomes marginal.

Static configurations $\chi=\chi(x)$ have energy
\begin{align}
E[\chi]
&= \int dx\left[\frac{1}{2}\big(\chi'(x)\big)^2 + V_{\rm eff}(\chi(x))\right],
\end{align}
and satisfy the Euler--Lagrange equation $\chi''(x)=dV_{\rm eff}/d\chi$. In the mechanical analogy, the quantity
\begin{align}
\mathcal E \equiv \frac{1}{2}\big(\chi'(x)\big)^2 - V_{\rm eff}(\chi(x))
\end{align}
is conserved along $x$. Finite-energy configurations interpolating between degenerate vacua must obey $\mathcal E=0$, which reduces the second-order equation to the first-order kink condition
\begin{align}
\frac{d\chi}{dx} &= \pm\sqrt{2\,V_{\rm eff}(\chi)}.
\label{eq:kink_first_order}
\end{align}
The corresponding topological sector is labeled by the winding charge
\begin{align}
Q \equiv \frac{\beta}{2\pi}\big[\chi(+\infty)-\chi(-\infty)\big]\in\mathbb{Z},
\label{eq:top_charge}
\end{align}
so that $Q=\pm 1$ describes a kink or antikink connecting neighboring vacua, for instance $\chi(-\infty)=0$ and $\chi(+\infty)=2\pi/\beta$.

When the logarithmic term in \eqref{eq:Veff_chi_explicit} is perturbatively small compared to the sine--Gordon piece, namely when $e^2K\beta^2/(4\pi m_\chi^2)\ll 1$, these kinks are smooth deformations of the standard sine--Gordon soliton. In the limit $e^2K\beta^2/(4\pi m_\chi^2)\to 0$ the usual profile
\begin{align}
\chi_0(x) &= \frac{4}{\beta}\arctan\!\big(e^{m_\chi(x-x_0)}\big)
\label{eq:SG_kink_profile}
\end{align}
is recovered, with tension $T_{\rm SG}=8m_\chi/\beta^2$.

\begin{figure}[t]
\centering
\includegraphics[width=\linewidth]{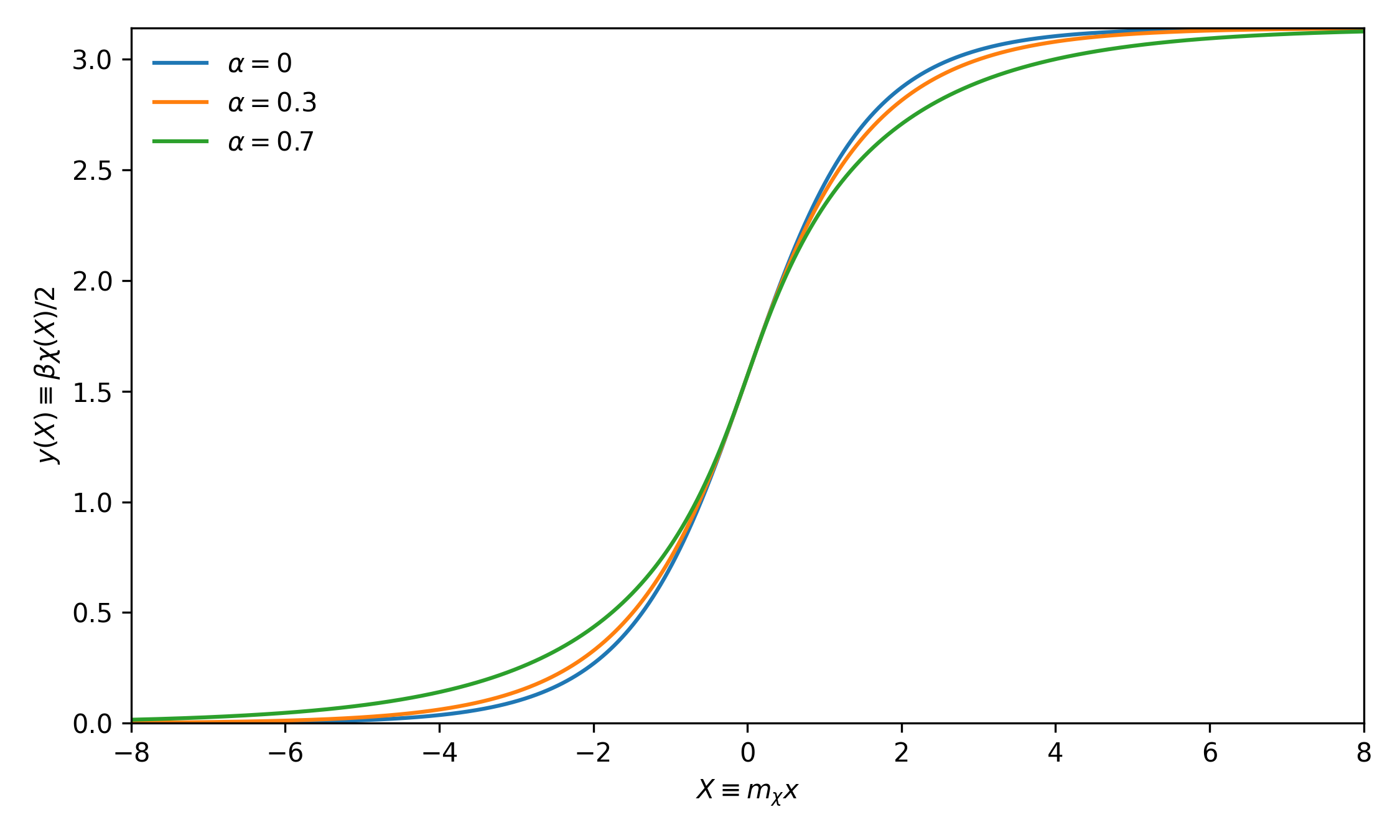}
\caption{\scriptsize Modified kink profiles in the one-loop sine--Gordon EFT. We plot \(y(X)\equiv \beta\chi(X)/2\) as a function of \(X\equiv m_\chi x\) for kinks interpolating between neighboring vacua \(y(-\infty)=0\) and \(y(+\infty)=\pi\). The \(\alpha=0\) curve reproduces the standard sine--Gordon profile \eqref{eq:SG_kink_profile}, while \(\alpha>0\) corresponds to integrating the first-order kink condition \eqref{eq:kink_first_order} with the one-loop effective potential associated with \eqref{eq:Leff_chi_final}.}
\label{fig:kink_profiles}
\end{figure}

A convenient way to quantify the deformation of the soliton sector induced by the logarithmic term is to compare kink profiles obtained from the first-order condition \eqref{eq:kink_first_order}. Working in the dimensionless coordinate \(X\equiv m_\chi x\) and plotting the rescaled field \(y(X)\equiv \beta\chi(X)/2\), Figure~\ref{fig:kink_profiles} contrasts the standard sine--Gordon kink (\(\alpha=0\)) with the profiles obtained by integrating \eqref{eq:kink_first_order} with the one-loop potential implied by \eqref{eq:Leff_chi_final}, for representative values of $\alpha$ within the stable single-field regime. The larger values should be read as a qualitative indication of the trend rather than as a strictly perturbative expansion.

Figure~\ref{fig:kink_profiles} makes this deformation concrete at the level of field profiles. All curves interpolate between the same neighboring vacua, but as $\alpha$ increases the interpolation is no longer the smooth sine--Gordon shape: the profile reacts to the loop-induced barrier and the transition region around $y\simeq\pi/2$ becomes increasingly localized. This is precisely the region where $m_\Psi(\chi)$ is smallest, so the figure should be viewed as a faithful description of the tails (where the EFT is controlled) and as indicative of the trend in the core as one approaches the regime where the integrated mode must be reinstated.

A leading estimate of the quantum correction to the tension is obtained by evaluating the logarithmic contribution on the unperturbed kink background. Writing
\begin{align}
\delta V(\chi)\equiv \frac{e^2K}{2\pi}\,\cos^2\!\left(\frac{\beta\chi}{2}\right)\,
   \log\!\left[\cos^2\!\left(\frac{\beta\chi}{2}\right)\right],
\end{align}
one has, at leading order in the deformation,
\begin{align}
\Delta T
&\approx \int_{-\infty}^{+\infty} dx\,\delta V\big(\chi_0(x)\big)
 = \frac{e^2K}{2\pi}\int_{-\infty}^{+\infty} dx\,
\tanh^2\!\big(m_\chi(x-x_0)\big)\,
\log\!\Big[\tanh^2\!\big(m_\chi(x-x_0)\big)\Big],
\end{align}
or, with $X\equiv m_\chi(x-x_0)$,
\begin{align}
\Delta T
&\approx \frac{e^2K}{2\pi m_\chi}\int_{-\infty}^{+\infty} dX\,
\tanh^2 X\,\log\!\big(\tanh^2 X\big)
= \frac{e^2K}{2\pi m_\chi}\left(4-\frac{\pi^2}{2}\right).
\end{align}
The integral evaluates in closed form: setting $u=\tanh X$ and using the series expansion of $(1-u^2)^{-1}$ on $0<u<1$, one obtains the exact result
\begin{align}
\int_{-\infty}^{+\infty} dX\,\tanh^2 X\,\log(\tanh^2 X) = 4 - \frac{\pi^2}{2}.
\label{eq:tension_integral_exact}
\end{align}
Since $4-\pi^2/2 \approx -0.935 < 0$, the kink tension is indeed reduced, $\Delta T < 0$, consistently with the downward shift in the small-oscillation mass \eqref{eq:Mchi2}. This closed-form expression is a concrete analytic prediction of the one-loop EFT. Quantitatively, however, the kink core probes the region around $\chi\simeq\pi/\beta$, where $m_\Psi^2(\chi)\to 0$ by \eqref{eq:mPsi_of_chi} and the single-field EFT becomes marginal. For this reason, while \eqref{eq:Leff_chi_final} captures the topology and the qualitative deformation of the soliton sector, a controlled determination of the core profile and of the tension ultimately requires reinstating the reduced gauge mode $\Psi$ and analyzing the soliton in the full two-field theory.

Having established the two complementary single-field EFT limits, we now quantize the coupled low-mode sector under a single spatial-mode truncation along $x$.

\section{Effective Quantum Optics Models}\label{Sec:QO}

The single-field EFTs of the preceding sections carry concrete phenomenological signatures even before quantization. In the chiral sector, the cleanest handles are the renormalized small-oscillation gap $V_{\rm eff}''(\chi_\star)$ and the induced deformation of kink profiles: tunnel-coupled one-dimensional superfluids, where the relative phase obeys an effective sine--Gordon dynamics, provide a direct cold-atom analogue in which these features are accessible through matter-wave interferometry, equal-time correlations, and non-equilibrium quenches \cite{Gritsev2007,DallaTorre2013,Schweigler2017,Pigneur2018}. In the gauge sector, superconducting circuits and Josephson networks---reviewed from both circuit-QED and microwave-photonics perspectives \cite{Blais2021,Gu2017MicrowavePhotonics}---offer a complementary platform where mode frequency shifts, anharmonicities, and multi-photon nonlinear response can be measured with high precision and compared to the loop-corrected EFT prediction \cite{Nigg2012,Astafiev2012,Kirchmair2013,Boyd2020}.

The most direct operator-level diagnostics, however, emerge from the effective Hamiltonians obtained after quantizing the coupled low-mode sector and applying a single spatial-mode truncation along $x$. These Hamiltonians sit squarely within the toolbox of nonlinear cavity and circuit QED; their operator content and selection rules sharply distinguish the trivial and locally condensed branches without further assumptions about the large-field completion. We develop them in the two subsections below.

\subsection{TLS Reduction and Longitudinal Two-Photon Squeezing Limit}\label{subsec:QO_TLS_TPRM}

We start from the two-field \(1+1\) fluctuation theory in \eqref{CM11} and reorganize the potential so that all the \(\chi\)--dependence is carried by \(\cos(\beta\chi)\). This yields
\begin{align}
V(\chi,\Psi)
&=
\big(e^2 K\,\Psi^2 + 4K p^2 \pi^2\big)
+ \big(e^2 K\,\Psi^2 - 4K p^2 \pi^2\big)\cos(\beta\chi).
\label{eq:potentialCosBetaChi}
\end{align}
The first parenthesis contains the purely photonic quadratic term together with an \(x\)--independent constant contribution, while the second parenthesis encodes the nontrivial light-matter structure.

The canonical momenta are \(\Pi_\chi=\dot\chi\) and \(\Pi_\Psi=\dot\Psi\), so the Hamiltonian density reads
\begin{align}
\mathcal{H}(x,t)
&=
\frac12\,\Pi_\chi^2
+ \frac12\,\Pi_\Psi^2
+ \frac12\,(\partial_x \chi)^2
+ \frac12\,(\partial_x \Psi)^2
+ V(\chi,\Psi),
\label{eq:twofieldHamiltonianDensity}
\end{align}
with \(V(\chi,\Psi)\) given in \eqref{eq:potentialCosBetaChi}.

To make contact with a cavity/circuit QED description, we approximate each field by a single spatial mode. 
Operationally, this corresponds to working in a regime where the dynamics is dominated by one cavity eigenmode and the population of higher modes remains negligible, as is typical in weakly excited nonlinear-optics settings; see \cite{Boyd2020} and references therein. 
For the photonic sector we choose a real mode function $u(x)$ normalized as
\begin{align}
\int_0^{L_x}\!dx\,u^2(x)=1,
\qquad
-\partial_x^2 u(x)=k^2 u(x),
\end{align}
so that the boundary conditions at $x=0$ and $x=L_x$ merely discretize the allowed values of $k$ and play no further role in the operator reduction below. 
We then introduce a canonical mode coordinate $q_\Psi(t)$ through
\begin{align}
\Psi(x,t)=q_\Psi(t)\,u(x),
\qquad
\Pi_\Psi(x,t)=\dot q_\Psi(t)\,u(x).
\label{eq:Psi_mode_q}
\end{align}
For the chiral sector we keep the lowest (spatially homogeneous) mode,
\begin{align}
u_0(x)\equiv L_x^{-1/2},
\qquad
\int_0^{L_x}\!dx\,u_0^2(x)=1,
\end{align}
and define its canonical coordinate \(q_\chi(t)\) as
\begin{align}
\chi(x,t)=q_\chi(t)\,u_0(x)=\frac{q_\chi(t)}{\sqrt{L_x}},
\qquad
\Pi_\chi(x,t)=\dot q_\chi(t)\,u_0(x)=\frac{\dot q_\chi(t)}{\sqrt{L_x}}.
\label{eq:chi_mode_q}
\end{align}
With this truncation the periodic operator becomes
\begin{align}
\beta\,\chi(x,t)=\beta\,\frac{q_\chi(t)}{\sqrt{L_x}}.
\end{align}

Substituting \eqref{eq:Psi_mode_q} and \eqref{eq:chi_mode_q} into \(\int_0^{L_x}dx\,\mathcal H\) and using the eigenvalue equation for \(u(x)\) yields a reduced Hamiltonian
\begin{align}
H \;=\; H_\Psi^{(0)} + H_\chi^{(0)} + H_{\rm int} + \text{(constant)}.
\end{align}
It is useful to display first the potential part of the single-mode Hamiltonian before splitting it into free and interaction pieces. Defining
\begin{align}
C(q_\chi)
\equiv
\cos\!\Big(\beta\,\frac{q_\chi}{\sqrt{L_x}}\Big),
\end{align}
one obtains
\begin{align}
V_{\rm red}(q_\chi,q_\Psi)
&=
e^2K\,q_\Psi^2\bigl(1+C(q_\chi)\bigr)
+4Kp^2\pi^2L_x\bigl(1-C(q_\chi)\bigr).
\label{eq:Vred_single_mode}
\end{align}
The factor \(L_x\) multiplying the purely chiral term follows from the spatially homogeneous chiral mode, whereas the \(\Psi^2\) term is weighted by the normalized cavity profile \(\int_0^{L_x}u^2dx=1\).

The free photonic part is
\begin{align}
H_\Psi^{(0)}
&=\frac12\,p_\Psi^2+\frac12\,\omega^2\,q_\Psi^2,
\qquad
p_\Psi\equiv \dot q_\Psi,
\qquad
\omega^2\equiv k^2+m_\Psi^2,
\label{eq:H0_Psi_q}
\end{align}
where \(m_\Psi^2=4e^2K\) as in \eqref{masses}. The chiral part is naturally taken as the full single-mode periodic potential,
\begin{align}
H_\chi^{(0)}
&=
\frac12\,p_\chi^2
+4Kp^2\pi^2L_x
\left[1-
\cos\!\Big(\beta\,\frac{q_\chi}{\sqrt{L_x}}\Big)
\right],
\qquad
p_\chi\equiv \dot q_\chi.
\label{eq:H0_chi_q}
\end{align}
Near the chosen minimum it is harmonic,
\begin{align}
H_\chi^{(0)}
&=
\frac12\,p_\chi^2+\frac12\,\Omega_\chi^2\,q_\chi^2+\ldots,
\qquad
\Omega_\chi^2\equiv m_\chi^2,
\end{align}
where the dots denote the anharmonic corrections inherited from the periodic potential.

The interaction term is therefore the residual photon--chiral coupling left after the quadratic photonic curvature at \(\chi=0\) has been included in \(H_\Psi^{(0)}\),
\begin{align}
H_{\rm int}
&=
e^2K\,q_\Psi^2
\left[
\cos\!\Big(\beta\,\frac{q_\chi}{\sqrt{L_x}}\Big)-1
\right].
\label{eq:Hint_q}
\end{align}

We canonically quantize the truncated coordinates by introducing ladder operators \(a,a^\dagger\) and \(b,b^\dagger\),
\begin{align}
q_\Psi = \frac{1}{\sqrt{2\omega}}\,(a+a^\dagger),
\qquad
p_\Psi = -\,i\sqrt{\frac{\omega}{2}}\,(a-a^\dagger),
\qquad [a,a^\dagger]=1,
\label{eq:qPsi_quant}
\\[4pt]
q_\chi = \frac{1}{\sqrt{2\Omega_\chi}}\,(b+b^\dagger),
\qquad
p_\chi = -\,i\sqrt{\frac{\Omega_\chi}{2}}\,(b-b^\dagger),
\qquad [b,b^\dagger]=1.
\label{eq:qChi_quant}
\end{align}
Then
\begin{align}
H_\Psi^{(0)}=\omega\left(a^\dagger a+\frac12\right),
\label{eq:singleModePhotonHamiltonian}
\end{align}
and the quadratic part of the chiral Hamiltonian is \(\Omega_\chi\left(b^\dagger b+\dfrac12\right)\).

In the low-energy regime we retain the two lowest chiral eigenstates within a single well,
\begin{align}
|g\rangle\equiv|0\rangle,
\qquad
|e\rangle\equiv|1\rangle,
\qquad
E_e-E_g = \Omega_\chi,
\end{align}
so that, after subtracting \((E_e+E_g)/2\),
\begin{align}
H_\chi^{(0)}\ \longrightarrow\ \frac{\Omega_\chi}{2}\,\sigma_z.
\end{align}

The operator that mediates the coupling to the photon mode is
\begin{align}
\hat C \;\equiv\; \cos\!\Big(\beta\,\frac{q_\chi}{\sqrt{L_x}}\Big).
\end{align}
Projected onto the TLS, it admits the Pauli decomposition
\begin{align}
\hat C \ \to\
C_0\,\mathds{1}+C_x\,\sigma_x+C_y\,\sigma_y+C_z\,\sigma_z.
\end{align}
By fixing the phases of \(|g\rangle\) and \(|e\rangle\) one can take the relevant matrix elements to be real, hence \(C_y=0\).

If the local chiral dynamics around the chosen minimum is parity-symmetric under \({q_\chi\to -q_\chi}\), the eigenstates \(|g\rangle\) and \(|e\rangle\) can be chosen with definite parity, while \(\hat C\) is parity-even. In that case the off-diagonal matrix element vanishes,
\begin{align}
C_x=\langle e|\hat C|g\rangle = 0,
\label{eq:Cx_zero_parity}
\end{align}
and the two-photon coupling is purely longitudinal. A transverse channel can be activated by parity-breaking bias in the chiral sector (or by working in a TLS basis that mixes opposite parities), which would yield \(C_x\neq 0\) and generate a term \(\propto\sigma_x(a^2+a^{\dagger 2})\). In the present paper we focus on the symmetry-preserving case \eqref{eq:Cx_zero_parity}.

For later reference, we record the leading small-fluctuation estimates in the parity-symmetric regime. Expanding \(\hat C\) to quadratic order in \(q_\chi\) and using \(\langle 0|q_\chi^2|0\rangle=\dfrac{1}{2\Omega_\chi}\) and \({\langle 1|q_\chi^2|1\rangle=\dfrac{3}{2\Omega_\chi}}\), one finds
\begin{align}
\langle g|\hat C|g\rangle
&\simeq 1-\frac{\beta^2}{4L_x\,\Omega_\chi},
\qquad
\langle e|\hat C|e\rangle
\simeq 1-\frac{3\beta^2}{4L_x\,\Omega_\chi},
\\[4pt]
C_0&\simeq 1-\frac{\beta^2}{2L_x\,\Omega_\chi},
\qquad
C_z\simeq -\,\frac{\beta^2}{4L_x\,\Omega_\chi},
\qquad
C_x=0.
\label{eq:C0Cz_estimates}
\end{align}

Quantizing \eqref{eq:Hint_q} with \eqref{eq:qPsi_quant} and projecting \(\hat C\) onto the TLS gives
\begin{align}
H_{\rm int}
&=
e^2K\,\frac{1}{2\omega}\,(a+a^\dagger)^2
\Big[(C_0-1)\,\mathds{1}+C_z\,\sigma_z\Big],
\end{align}
where we used \(C_x=0\) from \eqref{eq:Cx_zero_parity}. Normal ordering yields \((a+a^\dagger)^2=a^2+a^{\dagger2}+2a^\dagger a+1\). The terms proportional to \((C_0-1)\mathds{1}\) are independent of the TLS state and define a common quadratic renormalization of the cavity oscillator, including both number-conserving and squeezing pieces. Equivalently, they can be absorbed by a state-independent Bogoliubov redefinition of the reference oscillator, together with the corresponding redefinition of \(\omega\). The \(a\)-independent part proportional to \(C_z\sigma_z\) renormalizes the TLS splitting \(\Omega_\chi\). We therefore keep explicit only the genuinely state-dependent quadratic photon processes controlled by \(C_z\).

With this convention, the effective Hamiltonian in the parity-symmetric regime can be written as
\begin{align}
H_{\rm eff}
&\simeq
\omega\,a^\dagger a
+ \frac{\Omega_\chi}{2}\,\sigma_z
+ g_z\,\sigma_z\big(a^2+a^{\dagger2}+2a^\dagger a\big),
\label{eq:HeffRabiLike}
\end{align}
with the longitudinal two-photon coupling
\begin{align}
g_z \equiv e^2K\,\frac{1}{2\omega}\,C_z.
\end{align}
Using \eqref{eq:C0Cz_estimates}, one obtains the leading estimate
\begin{align}
g_z \simeq -\,\frac{e^2K}{2\omega}\,\frac{\beta^2}{4L_x\,\Omega_\chi}.
\end{align}
Equation \eqref{eq:HeffRabiLike} makes explicit that, within the symmetry-preserving truncation dictated by the underlying chiral dynamics, the dominant light-matter interaction is longitudinal and quadratic in the cavity field. It combines a dispersive shift \(2g_z\,\sigma_z\,a^\dagger a\) with a state-dependent squeezing channel \(g_z\,\sigma_z(a^2+a^{\dagger2})\). Any genuine transverse two-photon exchange term requires \(C_x\neq 0\), which is excluded by \eqref{eq:Cx_zero_parity} but can be engineered by controlled departures from the symmetry point.

\begin{figure}[t]
\centering
\includegraphics[width=0.98\textwidth]{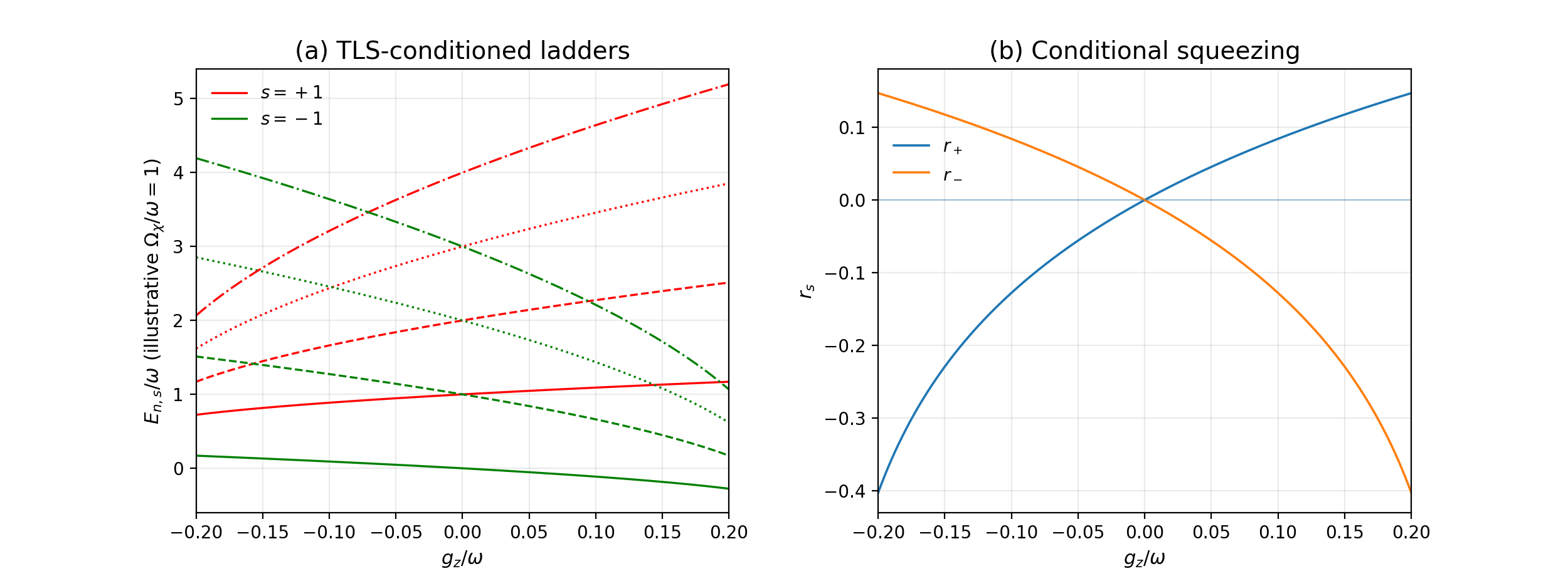}
\caption{\footnotesize
TLS-conditioned quadratic photon dynamics in the parity-symmetric longitudinal two-photon squeezing limit \eqref{eq:HeffRabiLike}.
Panel (a) shows the low-lying ladders \(E_{n,s}\) for \(n=0,1,2,3\) as a function of \(g_z/\omega\) in the two TLS sectors \(\sigma_z=s=\pm1\), with dressed frequency \(\omega_s\) given in \eqref{eq:rs_omegas}.
Panel (b) shows the corresponding squeezing parameter \(r_s\) from \eqref{eq:rs_omegas}.
The stability boundaries occur at \(g_z/\omega=\pm1/4\) (outside the plotted range).
}
\label{fig:QO_TPRM}
\end{figure}

To connect directly with standard quantum-optics diagnostics, we diagonalize the photonic sector conditioned on a fixed TLS eigenvalue $\sigma_z=s=\pm1$. Up to an irrelevant constant, \eqref{eq:HeffRabiLike} reduces to the quadratic bosonic Hamiltonian
\begin{align}
H^{(s)}_{\rm ph}
&=
A_s\,a^\dagger a+\frac{B_s}{2}\,\big(a^2+a^{\dagger2}\big),
\qquad
A_s\equiv \omega+2s g_z,
\qquad
B_s\equiv 2s g_z,
\end{align}
which is brought to diagonal form by a Bogoliubov transformation $a=\cosh r_s\,b_s-\sinh r_s\,b_s^\dagger$. The corresponding squeezing parameter and dressed oscillator frequency are
\begin{align}
r_s
&=
\frac14\log\!\left(\frac{A_s+B_s}{A_s-B_s}\right),
\qquad
\omega_s
=
\sqrt{A_s^{\,2}-B_s^{\,2}}
=
\sqrt{\omega^2+4\omega\,s g_z}.
\label{eq:rs_omegas}
\end{align}
Stability requires $A_s>|B_s|$, i.e.\ $\omega_s^2>0$, which translates into the simple bounds $g_z/\omega>-1/4$ for $s=+1$ and $g_z/\omega<1/4$ for $s=-1$. Figure~\ref{fig:QO_TPRM} illustrates how the TLS state controls both the level spacings and the amount of squeezing in the parity-symmetric regime.\footnote{For the illustrative ladder plot we set $\Omega_\chi/\omega=1$; the qualitative conditioning of the spectrum and of $r_s$ is independent of this choice.}

To guide the reading of Figure~\ref{fig:QO_TPRM}, note first that the conditional diagonalization implies that, within a fixed TLS branch $s=\pm1$, the entire bosonic ladder is organized by the single dressed frequency $\omega_s$ in \eqref{eq:rs_omegas}: the spacing between successive levels is $\omega_s$ and the overall offset is set by $\pm\Omega_\chi/2$. Panel (a) therefore consists of two families of approximately harmonic ladders, one for each TLS eigenvalue, with the four line styles corresponding to the lowest excitations $n=0,1,2,3$. As $g_z$ is tuned, the $s=+1$ branch hardens ($\omega_{+}$ increases) while the $s=-1$ branch softens ($\omega_{-}$ decreases), so that the level spacings become increasingly TLS-selective even though the interaction remains purely longitudinal and parity symmetric.

Panel (b) encodes the same physics in a genuinely quantum-optical language: the quadratic term $\propto\sigma_z(a^2+a^{\dagger2})$ prepares, within each TLS sector, a squeezed vacuum with squeezing parameter $r_s$. In terms of the standard quadratures $X\equiv (a+a^{\dagger})/\sqrt{2}$ and $P\equiv i(a^{\dagger}-a)/\sqrt{2}$, the Bogoliubov ground state has variances $\langle(\Delta X)^2\rangle_s\propto e^{-2r_s}$ and $\langle(\Delta P)^2\rangle_s\propto e^{+2r_s}$, so the sign of $r_s$ specifies which quadrature is squeezed. The rapid growth of $|r_s|$ upon approaching the stability bounds at $g_z/\omega=\pm1/4$ is the direct counterpart of the softening $\omega_s\to0$ in \eqref{eq:rs_omegas}.

The effective single-mode Hamiltonian derived above can be implemented directly in parametrically engineered resonator-qubit architectures \cite{Blais2021,Gu2017MicrowavePhotonics,Nigg2012,Chen2012}, where the longitudinal two-photon interaction can be characterized spectroscopically through photon-number-resolved dispersive shifts, Kerr ladders, and squeezing signatures \cite{Kirchmair2013}. Controlled departures from the symmetry point provide a clean knob to activate transverse channels and access the genuine two-photon quantum Rabi regime \cite{FornDiaz2019,Kockum2019,Felicetti2018}. A key point of our construction is that the effective couplings are not introduced as ad hoc fit parameters: they are inherited from the underlying microscopic completion through well-defined mode-overlap data and quantization prescriptions, so experiments can in principle test correlated parameter relations rather than isolated measurements.

\newpage

To complement the TLS-based longitudinal two-photon squeezing limit above, we now return to the one-loop effective action for the reduced gauge mode and expand about its trivial and nontrivial extrema. This produces quartic single-mode Hamiltonians whose normally ordered form makes the associated selection rules and mixing channels manifest.

\subsection{Quartic Single-Mode Photonic Hamiltonians: Trivial vs.\ Locally Condensed Branch}
\label{subsec:5B_quartic_Psi}

We start from the one-loop effective Lagrangian for the lowest reduced gauge mode, \eqref{eq:LeffPsi_1loop},
which defines an effective potential $V_{\rm eff}(\Psi)$ through
$\mathscr{L}_{\rm eff} = -\frac12\,\partial_\mu\Psi\,\partial^\mu\Psi - V_{\rm eff}(\Psi)$.
As above, we use the dimensionless variable $\tau$ (introduced previously) so that $\tau\propto \Psi^2$. This makes manifest that the theory around the trivial branch is $\Psi\to-\Psi$ symmetric, hence only even powers appear in any finite truncation around $\Psi=0$.

\subsubsection{Expansion About the Trivial Branch \texorpdfstring{$\Psi=0$}{Psi = 0} and Normal-Ordered Hamiltonian}
\label{subsubsec:5B_trivial}

Expanding \eqref{eq:LeffPsi_1loop} for small $\tau\ll 1$ and truncating at $\mathcal{O}(\Psi^4)$, the quartic coupling evaluates to
\begin{align}
\lambda^{(0)}_4
=
-\,\frac{18\,M^2}{\pi}\left(\frac{e^2K\beta^2}{m_\chi^2}\right)^2.
\label{eq:lambda4_trivial}
\end{align}
The local EFT then reads
\begin{align}
\mathscr{L}_{\mathrm{eff}}[\Psi]
\simeq
-\frac{1}{2}\,\partial_\mu \Psi\,\partial^\mu \Psi
-\frac{1}{2}\,m_\Psi^2\,\Psi^2
-\frac{\lambda^{(0)}_4}{4!}\,\Psi^4,
\qquad
m_\Psi^2=4e^2K.
\label{eq:LeffPsi_trivial_quartic}
\end{align}
The sign in \eqref{eq:lambda4_trivial} is a concise way to encode an important point: at quartic order the interaction is attractive, and stability is only recovered once the higher-order terms (already present in \eqref{eq:LeffPsi_1loop}) are reinstated. For the purpose of a controlled weak-amplitude expansion, however, \eqref{eq:LeffPsi_trivial_quartic} is the correct local truncation.

We repeat the same single-mode reduction used in Subsection~\ref{subsec:QO_TLS_TPRM}, i.e.\ we insert \eqref{eq:Psi_mode_q} into the spatial integral of \eqref{eq:LeffPsi_trivial_quartic}. Denoting the mode overlap
\begin{align}
I_n \equiv \int_0^{L_x}\!dx\,u^n(x),
\label{eq:In_def}
\end{align}
the reduced Hamiltonian takes the standard anharmonic oscillator form
\begin{align}
H^{(0)}_\Psi
=
\frac12\,p_\Psi^2+\frac12\,\omega^2\,q_\Psi^2
+\frac{\lambda^{(0)}_4}{4!}\,I_4\,q_\Psi^4,
\qquad
\omega^2=k^2+m_\Psi^2,
\end{align}
where $\omega$ is the same oscillator frequency defined in \eqref{eq:H0_Psi_q}.
Quantizing with \eqref{eq:qPsi_quant} and writing the result in fully normal-ordered form yields
\begin{align}
H^{(0)}_\Psi
&=
\omega\,a^\dagger a
+ \delta\omega_0\,a^\dagger a
+ g^{(0)}_{2}\,\big(a^2+a^{\dagger 2}\big)
+ g^{(0)}_{31}\,\big(a^{\dagger 3}a+a^\dagger a^3\big)
+ g^{(0)}_{\rm K}\,a^{\dagger 2}a^2
+ g^{(0)}_{4}\,\big(a^4+a^{\dagger 4}\big)
+E^{(0)}_{\rm vac},
\label{eq:H_trivial_normalordered}
\end{align}
with couplings fixed by a single microscopic parameter,
\begin{align}
\kappa_0 \equiv \frac{\lambda^{(0)}_4\,I_4}{96\,\omega^2},
\label{eq:kappa0_def}
\end{align}
as
\begin{align}
\delta\omega_0 = 12\,\kappa_0,
\qquad
g^{(0)}_{2}=6\,\kappa_0,
\qquad
g^{(0)}_{31}=4\,\kappa_0,
\qquad
g^{(0)}_{\rm K}=6\,\kappa_0,
\qquad
g^{(0)}_{4}=\kappa_0,
\qquad
E^{(0)}_{\rm vac}=\frac{\omega}{2}+3\,\kappa_0.
\label{eq:couplings_trivial}
\end{align}
Two comments make \eqref{eq:H_trivial_normalordered} more transparent. First, normal ordering has turned the single local operator $q_\Psi^4$ into a tower of number-conserving and nonconserving channels: besides the Kerr term $a^{\dagger 2}a^2$ one obtains a two-photon squeezing piece $a^2+a^{\dagger 2}$, as well as number-nonconserving four-photon processes $a^4+a^{\dagger 4}$ and the mixed $a^{\dagger 3}a+a^\dagger a^3$. Second, the same normal ordering also produces the quadratic renormalizations $\delta\omega_0$ and $g^{(0)}_2$. If desired, one can absorb $\delta\omega_0$ into a redefinition of $\omega$ (as already done conceptually in Subsection~\ref{subsec:QO_TLS_TPRM} for dispersive shifts); we keep it explicit here because it is fixed by the same $\kappa_0$ that controls the genuine nonlinearities.

\subsubsection{Expansion About a Locally Condensed Branch \texorpdfstring{$\Psi_\star\neq 0$}{Psi* != 0} and Normal-Ordered Hamiltonian}
\label{subsubsec:5B_nontrivial}

We expand around the locally condensed branch discussed above, $\Psi_\star\neq 0$, characterized by the corresponding $\tau_\star$ (fixed there by the gap equation \eqref{eq:gap-photon}). This assumes that the local branch belongs to the retained longitudinal low mode. If the boundary conditions require a nontrivial cavity profile, $\Psi_\star$ should be interpreted as the amplitude of that retained low mode rather than as a literal constant four-dimensional gauge field. We write
\begin{align}
\Psi(x,t)=\Psi_\star+\xi(x,t),
\label{eq:Psi_shift_xi}
\end{align}
and construct the local expansion of the effective potential around $\Psi_\star$ up to quartic order in $\xi$,
\begin{align}
\mathscr{L}_{\mathrm{eff}}[\xi]
\simeq
-\frac{1}{2}\,\partial_\mu \xi\,\partial^\mu \xi
-\frac{1}{2}\,m_\star^2\,\xi^2
-\frac{\lambda^{(\star)}_3}{3!}\,\xi^3
-\frac{\lambda^{(\star)}_4}{4!}\,\xi^4,
\label{eq:Leff_xi_quartic}
\end{align}
where $m_\star^2\equiv V_{\rm eff}''(\Psi_\star)$ and $\lambda^{(\star)}_{3,4}\equiv V_{\rm eff}^{(3,4)}(\Psi_\star)$.
In terms of $\tau_\star$ and the shorthand
\begin{align}
L_\star \equiv \log\!\big[(\tau_\star-1)^2\big],
\end{align}
one convenient parametrization of these coefficients is
\begin{align}
m_\star^2
&=
-\,\frac{2M^2}{\pi}\left(\frac{e^2K\beta^2}{m_\chi^2}\right)\,\tau_\star\,(L_\star+3),
\label{eq:mstar2_def}
\\[4pt]
\lambda^{(\star)}_3
&=
-\,\frac{2M^2}{\pi}\left(\frac{e^2K\beta^2}{m_\chi^2}\right)^{2}\,
\Psi_\star\,
\frac{(3L_\star+13)\tau_\star-3L_\star-9}{\tau_\star-1},
\\[4pt]
\lambda^{(\star)}_4
&=
-\,\frac{2M^2}{\pi}\left(\frac{e^2K\beta^2}{m_\chi^2}\right)^{2}\,
\frac{
3L_\star\,\tau_\star^2-6L_\star\,\tau_\star+3L_\star
+25\tau_\star^2-42\tau_\star+9
}{(\tau_\star-1)^2}.
\end{align}
The qualitative difference with the trivial branch is immediate already at the level of the local EFT \eqref{eq:Leff_xi_quartic}: the $\Psi\to-\Psi$ symmetry is broken once we expand about $\Psi_\star\neq 0$, and therefore odd operators (starting with $\xi^3$) are allowed and generically present.

We again reduce to the same single spatial mode as in Subsection~\ref{subsec:QO_TLS_TPRM}, taking $\xi(x,t)=q_\xi(t)\,u(x)$ with the same normalized profile $u(x)$ used in \eqref{eq:Psi_mode_q}. Using \eqref{eq:In_def}, the reduced Hamiltonian becomes
\begin{align}
H^{(\star)}_\xi
=
\frac12\,p_\xi^2+\frac12\,\omega_\star^2\,q_\xi^2
+\frac{\lambda^{(\star)}_3}{3!}\,I_3\,q_\xi^3
+\frac{\lambda^{(\star)}_4}{4!}\,I_4\,q_\xi^4,
\qquad
\omega_\star^2 = k^2+m_\star^2.
\label{eq:Hq_star}
\end{align}
Quantizing as in \eqref{eq:qPsi_quant} (now with $\omega\to\omega_\star$ and $q_\Psi\to q_\xi$) and writing the result in fully normal-ordered form gives
\begin{align}
H^{(\star)}_\xi
&=
\omega_\star\,a^\dagger a
+ \kappa_3\Big[
a^{\dagger 3}+3a^{\dagger 2}a+3a^\dagger a^2+a^3
+3\big(a^\dagger+a\big)
\Big]
\nonumber\\
&\quad
+ \kappa_4\Big[
a^{\dagger 4}+4a^{\dagger 3}a+6a^{\dagger 2}a^2+4a^\dagger a^3+a^4
+6\big(a^{\dagger 2}+2a^\dagger a+a^2\big)
\Big]
+E^{(\star)}_{\rm vac},
\label{eq:H_star_normalordered}
\end{align}
with the compact definitions
\begin{align}
\kappa_3 \equiv \frac{\lambda^{(\star)}_3\,I_3}{3!\,(2\omega_\star)^{3/2}},
\qquad
\kappa_4 \equiv \frac{\lambda^{(\star)}_4\,I_4}{96\,\omega_\star^2},
\qquad
E^{(\star)}_{\rm vac}\equiv \frac{\omega_\star}{2}+3\,\kappa_4.
\end{align}
Equation \eqref{eq:H_star_normalordered} makes the physics of the locally condensed expansion point particularly explicit. Compared with \eqref{eq:H_trivial_normalordered}, the new ingredient is the cubic coupling $\kappa_3$, which activates (i) genuine three-photon processes ($a^3+a^{\dagger 3}$ and the mixed $a^{\dagger 2}a+a^\dagger a^2$) and (ii) a linear channel $a+a^\dagger$ once the Hamiltonian is written in normal-ordered form. The latter is the single-mode avatar of the statement that, after expanding around the shifted branch, the natural local oscillator state need not be centered at the origin of the \emph{free} oscillator basis: a coherent displacement is generically induced, and one can remove the linear term by a further (small) canonical shift if one wishes to define ladder operators adapted to the interacting minimum. For the present goal---a side-by-side comparison of the operator content in the two branches---it is actually more informative to keep the normal-ordered Hamiltonian as in \eqref{eq:H_star_normalordered}, since it cleanly highlights the loss of the $\Psi\to-\Psi$ selection rule.

At this quartic level, the contrast between the two scenarios is sharp and operational. In the trivial branch the leading nonlinearity is even and all number-nonconserving channels originate from the single quartic parameter $\kappa_0$ in \eqref{eq:kappa0_def}; correspondingly, nonlinear response is constrained by a remnant parity symmetry, and odd-harmonic generation is absent in the strict single-mode truncation. In the nontrivial branch, already the cubic term (controlled by $\kappa_3$) predicts qualitatively new signatures: asymmetry under $\Psi\to-\Psi$, activation of three-photon mixing channels, and a ground state that is naturally of squeezed-coherent type rather than purely squeezed. From the viewpoint of branch structure, the parameter regime where $\tau_\star$ approaches the critical value discussed earlier corresponds to a softening of $m_\star^2$ in \eqref{eq:mstar2_def} and hence of $\omega_\star$ in \eqref{eq:Hq_star}; this is the regime where the single-mode truncation becomes most sensitive to fluctuations and where the clean separation between ``background'' and ``fluctuation'' implicit in \eqref{eq:Psi_shift_xi} begins to degrade, as anticipated in the discussion of the local-condensation window. In that sense, \eqref{eq:H_star_normalordered} should be read as the controlled operator-level prediction away from the immediate critical neighborhood, while its progressive breakdown as $\omega_\star\to 0$ is itself a diagnostic of the approaching critical reorganization of the reduced EFT.

The contrast between the two branches also has clear experimental and phenomenological signatures. In the symmetry-preserving branch, the normally ordered single-mode Hamiltonian \eqref{eq:H_trivial_normalordered} is that of a Kerr-type nonlinear resonator supplemented by number-nonconserving quartic processes ($\Delta N=\pm 2,\pm 4$), i.e.\ the single-mode reduction of the familiar $\chi^{(3)}$ nonlinearity. Kerr Hamiltonians of this kind are routinely realized in circuit QED and characterized through photon-number-resolved Kerr ladders and collapse-and-revival dynamics; they also underpin the generation and stabilization of photonic cat states via engineered two-photon processes \cite{Kirchmair2013,Vlastakis2013,Leghtas2015,Mirrahimi2014}.

By contrast, the locally condensed branch \eqref{eq:H_star_normalordered} generically contains a leading cubic nonlinearity. After a coherent displacement---or in the presence of a strong pump tone---this term acts as an effective three-wave mixer for the residual fluctuations, the same resource exploited in Josephson three-wave-mixing devices for near-quantum-limited amplification, frequency conversion, and non-reciprocal microwave routing \cite{Bergeal2010,Frattini2017,Sliwa2015}. Conceptually, it plays the role of a $\chi^{(2)}$ nonlinearity in nonlinear optics, available only once the relevant inversion/parity symmetry is broken \cite{Kippenberg2011,Gaeta2019,Boyd2020}. From this viewpoint, the cleanest discriminator between the trivial and locally condensed branches is whether oscillator-parity is an autonomous selection rule: observing odd-$\Delta N$ mixing channels (or, equivalently, three-wave processes that cannot arise from a purely even potential) points to the locally condensed effective branch, whereas their absence is consistent with the trivial branch.

\begin{figure}[t]
\centering
\includegraphics[width=0.98\textwidth]{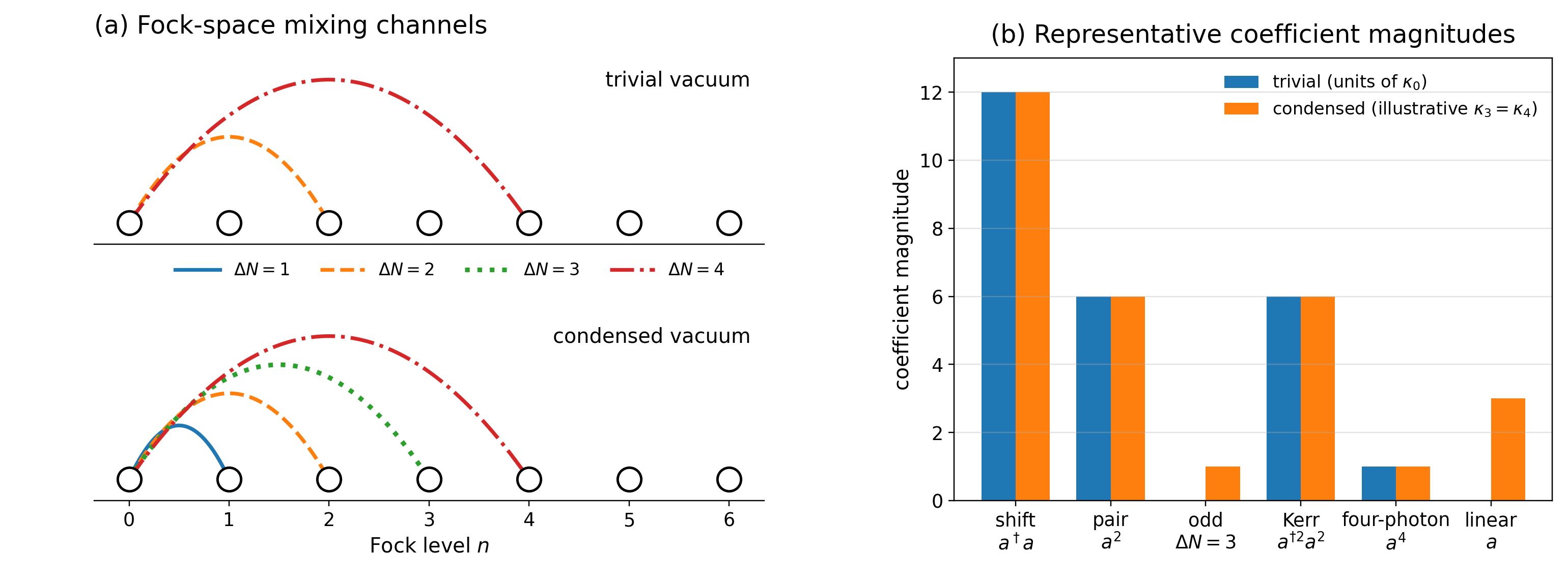}
\caption{\footnotesize
Quartic single-mode photonic Hamiltonians in the trivial and locally condensed branches.
Panel (a) visualizes the selection rules implied by the normally ordered Hamiltonians \eqref{eq:H_trivial_normalordered} and \eqref{eq:H_star_normalordered} in a truncated Fock basis: the trivial branch only connects levels with even photon-number difference (even $\Delta N$), while the locally condensed branch generically activates odd-$\Delta N$ channels through the cubic sector proportional to $\kappa_3$.
Panel (b) compares representative coefficient magnitudes for the leading operator families (number shift, $\Delta N=2$ pair processes, Kerr, $\Delta N=4$ processes, and the odd channels present only in the locally condensed branch). Trivial couplings are shown in units of $\kappa_0$ fixed by \eqref{eq:couplings_trivial}; condensed coefficients are displayed for $\kappa_3=\kappa_4$ to illustrate the relative weight of odd and even processes.
}
\label{fig:QO_quartic}
\end{figure}

To make this selection rule and the operator content of the two effective Hamiltonians fully explicit, we work directly in the Fock basis of the single mode. In the trivial branch \eqref{eq:H_trivial_normalordered} every interaction changes the photon number by an even integer, so the oscillator-parity operator $\Pi\equiv e^{i\pi a^\dagger a}$ is conserved and the Hilbert space decomposes into even and odd photon-number sectors that do not mix. In the locally condensed branch \eqref{eq:H_star_normalordered}, expanding around the displaced minimum activates odd operators through the $\kappa_3$ sector (in particular linear and cubic terms), so $\Pi$ is no longer conserved for the autonomous fluctuations and odd-$\Delta N$ transitions become allowed. Figure~\ref{fig:QO_quartic} summarizes this contrast: panel (a) visualizes the corresponding mixing channels in a truncated Fock basis, while panel (b) compares representative coefficient magnitudes (trivial couplings in units of $\kappa_0$ fixed by \eqref{eq:couplings_trivial}; condensed coefficients displayed for the illustrative choice $\eta\equiv\kappa_3/\kappa_4=1$, with the qualitative separation between even and odd channels persisting for generic $\eta$).

\newpage

\section{Discussions and Conclusions}\label{Sec:Conclusions}

This manuscript develops a finite-volume Chiral--Maxwell EFT in which a topologically nontrivial hadronic sector behaves as a nonlinear medium for a reduced cavity gauge mode. We start from leading-order Chiral Perturbation Theory minimally coupled to Maxwell theory, placed in a finite-volume geometry where baryon number is encoded as a topological charge. Forcing a nonzero baryonic charge into a finite region naturally favors spatially structured chiral configurations, as in other finite-density systems with inhomogeneous phases \cite{R1,R2,FuldeFerrell1964,LarkinOvchinnikov1965,pasta1}. Our main point is that such hadronic backgrounds can generate a gauge-invariant effective potential for the retained cavity coordinate and, within its controlled regime, a nontrivial locally stable reduced gauge-mode branch.

First, we construct a consistent and analytically tractable reduction to the lowest chiral and reduced gauge modes compatible with nontrivial finite-volume baryonic/topological sectors. The Ansatz of Section~\ref{Sec:Ansatz}, summarized by \eqref{Ansatz01} and \eqref{AnsatzGauge}, reduces the full $3+1$ Chiral--Maxwell dynamics to an effective \(1+1\) system while keeping the configuration genuinely three-dimensional in the topological sense encoded in \eqref{currents}. The geometric rationale is that transverse translations are realized equivariantly (symmetry up to internal/gauge transformations), so that gauge-invariant local densities are compatible with the Killing symmetries even though the fields themselves carry rigid winding. This viewpoint is standard in symmetric reductions and is spelled out in Appendix~\ref{App:LieAnsatz} \cite{ForgacsManton,ElgoodMeessenOrtin2020,CanforaMaeda2013,CanforaDiMauroKurkovNaddeo2015}. In the gauge sector, the cavity naturally promotes Wilson-line (holonomy) data on compact cycles to physical low-energy degrees of freedom; the relevant reduced gauge variable is therefore most naturally associated with the lowest holonomy mode(s) of this type rather than with a gauge-dependent local expectation value of $A_\mu$ \cite{Hosotani83,AitkenChermanPoppitzYaffe2017,Akamatsu2022}.

Second, we identify explicitly the regimes in which the coupled two-field \(1+1\) theory admits reliable single-mode effective descriptions. When the chiral sector is heavy relative to the reduced gauge coordinate, integrating it out at one loop yields an effective potential for the reduced gauge coordinate \(\Psi\) and allows one to delineate analytically the parameter window where nontrivial local minima appear. These minima are best interpreted as locally stable finite-volume reduced gauge-mode branches, or as the onset of a reduced-mode condensation instability, not as a statement about the global vacuum of the full theory after extrapolation to arbitrary field values. In the opposite hierarchy, integrating out the gauge mode yields a controlled one-loop corrected sine--Gordon-type EFT for the chiral mode, making explicit how the gauge sector reshapes the chiral landscape and clarifying where scale separation breaks down and the full coupled dynamics must be retained.

Third, we build a concrete bridge to effective quantum-optics models. The reduced dynamics organizes into a small set of cavity Hamiltonians (Kerr-type and related nonlinear oscillators) that are widely used in circuit QED, microwave photonics, and nonlinear optics \cite{Blais2021,Gu2017MicrowavePhotonics,Boyd2020}, and the two branches of the reduced theory correspond to qualitatively different selection rules and nonlinear processes. This provides a clear route to phenomenological diagnostics, in close analogy with the characterization of nonlinearities, multi-photon processes, and engineered dissipation in modern superconducting-cavity platforms \cite{Kirchmair2013,Vlastakis2013,Leghtas2015,Mirrahimi2014,Bergeal2010,Frattini2017,Sliwa2015,Kippenberg2011,Gaeta2019}.

The analysis also has clear limitations, which simultaneously point to natural extensions. We worked in a minimal ChPT truncation (although the subleading corrections can be included in a systematic way, as we will explore in future publications). We emphasized the lowest-mode sector selected by the cavity. The one-loop effective potential for the reduced gauge coordinate is a controlled object within the domain of validity identified in Sec.~\ref{subsec:validity}: the locally stable nontrivial branch lies inside this domain, and the apparent large-field descent of $V_{\mathrm{eff}}$ for $\tau \gg 1$ is an artefact of extrapolating beyond the regime where the integrated-out fermion is heavy, not a physical instability of the full theory. Near the boundaries of the local-condensation window, whenever the would-be heavy mode becomes parametrically light, or when one probes large values of the reduced coordinate beyond the controlled range of the one-loop potential, additional modes and higher-order EFT operators become relevant and the single-field EFTs are insufficient. These regimes are accessible within the same framework by keeping the coupled system and systematically enlarging the mode content.

Several future directions appear particularly promising. One is to include systematically the next operators in the ChPT expansion (pion mass term, higher-derivative terms, and, where relevant, anomalous contributions) and to assess how robust the local-condensation window is under these deformations \cite{ChPT1,ChPT2,ChPT3,ChPT4,ChPT5,ChPT6,ChPT7,ChPT8,Witten,skyrme}. Another direction is to study multi-mode generalizations in the cavity, including transverse excitations, to explore whether richer spatial patterns analogous to crystalline phases can coexist with (or compete against) the locally condensed reduced gauge-mode branch, building on the analytic finite-volume technology developed for Skyrme-type systems \cite{Fab0,Fab1,gaugsk,gaugsk2,crystal1,crystal2,crystal3,range1,range2,range3,range4}. Finally, adding driving and dissipation would connect the reduced cavity theory to recent open-EFT treatments of light in media and gauged degrees of freedom, where Schwinger--Keldysh/influence-functional methods encode noise while preserving gauge invariance \cite{SalcedoColasPajer2025OpenLight,KaplanekMylovaTolley2025OpenEFT}, and to the nonequilibrium setting of photon-condensation experiments \cite{photoncond1,photoncond2,photoncond3,photoncond4,photoncond5,photoncond6}.

\subsection*{Acknowledgements}
F. C. has been funded by FONDECYT Grants No. 1240048, 1240043 and 1240247 and also supported by Grant ANID EXPLORACIÓN 13250014. S.R. gratefully acknowledges support from Postdoctorado USS 2023 USS-FIN-23-PDOC-04 and FONDECYT Iniciación No. 11261586, as well as the hospitality and physical space provided by DFI-FCFM Universidad de Chile and CECs during the completion of this manuscript.

\subsection*{Statements and Declarations}
\textbf{Funding:} Details of financial support are given in the Acknowledgements section.

\textbf{Competing interests:} The authors declare no competing interests.

\textbf{Data availability:} No datasets were generated or analysed during the current study.

\appendix

\section{Lie-Derivative Rationale for the Cavity Ansatz}\label{App:LieAnsatz}

This appendix summarizes the geometric rationale for the cavity Ans\"atze of Section~\ref{Sec:Ansatz}, phrased in the compensated Lie-derivative language commonly used when spacetime isometries coexist with internal symmetries. The key idea is to realize translations along the compact transverse directions equivariantly: the fundamental fields may change by a rigid internal/gauge transformation, while all gauge-invariant local densities remain homogeneous. This is the standard mechanism behind symmetric/equivariant reductions in gauge theories \cite{ForgacsManton,ElgoodMeessenOrtin2020}.\footnote{Symmetry-compatible Ans\"atze in finite volume can be motivated in three standard ways: (i) equivariant reduction, i.e.\ invariance under isometries up to internal/gauge transformations \cite{ForgacsManton,ElgoodMeessenOrtin2020}; (ii) the weaker requirement that gauge-invariant local densities (in particular $T_{\mu\nu}$) inherit the isometries, $\mathcal L_K T_{\mu\nu}=0$, as in boson stars and stationary scalar hair \cite{Kaup1968,RuffiniBonazzola1969,SchunckMielke2003,LieblingPalenzuela2012,HerdeiroRadu2014}; (iii) an EFT/zero-mode viewpoint keeping only holonomies below the KK gap \cite{Hosotani83,AitkenChermanPoppitzYaffe2017,Akamatsu2022}.}

We work on the cavity background
\begin{align}
ds^{2}=-dt^{2}+dx^{2}+L^{2}\big(d\mathfrak y^{2}+d\mathfrak z^{2}\big),
\end{align}
with coordinate ranges $0\le x\le L_x$, $0\le \mathfrak y\le 2\pi$ and $0\le \mathfrak z\le 4\pi$. The transverse isometries are generated by the commuting Killing vectors
\begin{align}
K_{(\mathfrak y)}=\partial_{\mathfrak y},\qquad K_{(\mathfrak z)}=\partial_{\mathfrak z},\qquad [K_{(\mathfrak y)},K_{(\mathfrak z)}]=0.
\end{align}
We use the gauged chiral model defined in \eqref{eq:ChiralMaxwellAction}. For convenience we recall the basic building blocks,
\begin{align}
D_\mu U=\partial_\mu U + A_\mu [t_3,U],
\qquad
R_\mu=U^{-1}D_\mu U,
\end{align}
with gauge transformations
\begin{align}
U\to e^{\lambda t_3}Ue^{-\lambda t_3},
\qquad
A_\mu\to A_\mu-\partial_\mu\lambda.
\end{align}
Here and below we denote collectively by $x^\mu=(t,x,\mathfrak y,\mathfrak z)$ the full coordinate set, and we allow all fields to depend on all four coordinates prior to imposing the symmetry reduction.

For an ordinary tensor field $T$, strict homogeneity along a Killing direction $K$ is simply $\mathcal L_K T=0$. In gauge(-matter) systems this is too rigid: what one really wants is invariance of the action and of all gauge-invariant local observables, allowing the fundamental fields to shift by an internal/gauge transformation under the isometry. This is captured by compensated (or generalized) Lie derivatives \cite{ForgacsManton,ElgoodMeessenOrtin2020}.

For the sigma model field $U$, we allow left/right $\mathfrak{su}(2)$ compensators $\Lambda^{(L)}_K$ and $\Lambda^{(R)}_K$ and define
\begin{align}
\widehat{\mathcal L}_K U \equiv \mathcal L_K U-\Lambda^{(L)}_K\,U+U\,\Lambda^{(R)}_K,
\end{align}
so that equivariance means $\widehat{\mathcal L}_K U=0$. The compensators $\Lambda^{(L)}_K$ and $\Lambda^{(R)}_K$ are rigid internal (left/right) motions in the chiral $SU(2)$ target space used to realize the spacetime isometries equivariantly; they do not introduce any new local redundancy beyond the $U(1)$ gauging already specified. For an Abelian connection, ``invariant up to gauge'' means that the ordinary Lie derivative is pure gauge,
\begin{align}
\widehat{\mathcal L}_K A \equiv \mathcal L_K A-d\lambda_K,
\qquad
\widehat{\mathcal L}_K A=0,
\end{align}
with $\lambda_K$ the gauge parameter associated to the isometry generated by $K$. This condition allows residual dependence along the Killing direction that is pure gauge; after choosing a representative, the surviving gauge content along compact cycles is precisely the transverse zero-mode data (holonomies).

Equivariance also comes with an integrability requirement expressing the closure of the Killing algebra,
\begin{align}
[\widehat{\mathcal L}_{K_1},\widehat{\mathcal L}_{K_2}] = \widehat{\mathcal L}_{[K_1,K_2]}.
\end{align}
In the present cavity, $[\,\partial_{\mathfrak y},\partial_{\mathfrak z}\,]=0$, so one can consistently take constant commuting compensators (proportional to $t_3$) without obstruction.

The goal in the main text is a low-energy sector in which local gauge-invariant densities do not depend on $(\mathfrak y,\mathfrak z)$, while the configuration can still carry nontrivial winding data on the compact transverse cycles. The minimal way to do this is to realize translations in $\mathfrak y$ and $\mathfrak z$ as rigid motion along the $t_3$ direction, but arranged in different left/right channels so that the two cycles contribute independently. This logic is closely aligned with generalized hedgehog/equivariant constructions in Skyrme-type systems, especially in finite-volume settings where one wants nontrivial topological sectors without exciting full transverse dynamics \cite{CanforaMaeda2013,CanforaDiMauroKurkovNaddeo2015}.

Concretely, for $K_{(\mathfrak y)}=\partial_{\mathfrak y}$ choose
\begin{align}
\Lambda^{(L)}_{(\mathfrak y)}=0,
\qquad
\Lambda^{(R)}_{(\mathfrak y)}=p\,t_3,
\end{align}
so that $\widehat{\mathcal L}_{\partial_{\mathfrak y}}U=0$ becomes the first-order equation
\begin{align}
0=\widehat{\mathcal L}_{\partial_{\mathfrak y}}U=\partial_{\mathfrak y}U-U\,p\,t_3
\qquad\Longrightarrow\qquad
\partial_{\mathfrak y}U=U\,p\,t_3.
\end{align}
This integrates immediately to
\begin{align}
U(t,x,\mathfrak y,\mathfrak z)=\widetilde U(t,x,\mathfrak z)\,e^{t_3 p\mathfrak y}.
\end{align}
Similarly, for $K_{(\mathfrak z)}=\partial_{\mathfrak z}$ take
\begin{align}
\Lambda^{(L)}_{(\mathfrak z)}=p\,t_3,
\qquad
\Lambda^{(R)}_{(\mathfrak z)}=0,
\end{align}
so that
\begin{align}
0=\widehat{\mathcal L}_{\partial_{\mathfrak z}}U=\partial_{\mathfrak z}U-p\,t_3\,U
\qquad\Longrightarrow\qquad
\partial_{\mathfrak z}U=p\,t_3\,U,
\end{align}
and therefore
\begin{align}
U(t,x,\mathfrak y,\mathfrak z)=e^{t_3 p\mathfrak z}\,U_0(t,x)\,e^{t_3 p\mathfrak y}.
\end{align}
At this stage, the entire dependence on $(\mathfrak y,\mathfrak z)$ is fixed by symmetry: it is rigid group motion with a discrete label $p$. The remaining freedom is the $(t,x)$-dependent core $U_0(t,x)$, which is where the light dynamics lives.

To match the Euler-angle parametrization used in the main text,
\begin{align}
U=e^{t_3F}\,e^{t_2H}\,e^{t_3G},
\end{align}
we take the simplest one-parameter family for the core,
\begin{align}
U_0(t,x)=e^{t_2H(t,x)}.
\end{align}
This yields
\begin{align}
U=e^{t_3 p\mathfrak z}\,e^{t_2H(t,x)}\,e^{t_3 p\mathfrak y} \qquad\Longrightarrow\qquad
H=H(t,x),\ \  F=p\,\mathfrak z,\ \  G=p\,\mathfrak y,
\end{align}
which is precisely the chiral Ansatz stated in \eqref{EulerAnsatz} and \eqref{Ansatz01}. The integer $p$ labels the discrete winding sector compatible with the periodic identifications in $\mathfrak y$ and $\mathfrak z$ and with the requirement that $U$ be globally well-defined on the identified domain.

For the gauge field we impose the same notion of transverse symmetry, namely $\widehat{\mathcal L}_{\partial_{\mathfrak y}}A=\widehat{\mathcal L}_{\partial_{\mathfrak z}}A=0$. Taking constant $\lambda_{(\mathfrak y)}$ and $\lambda_{(\mathfrak z)}$ gives
\begin{align}
\partial_{\mathfrak y}A_\mu=\partial_{\mathfrak z}A_\mu=0,
\end{align}
so $A_\mu$ may depend on $(t,x)$ only. In the low-energy cavity regime, the relevant transverse gauge data are the zero-modes along the compact cycles, i.e.\ Wilson-line/holonomy degrees of freedom, which in general cannot be eliminated by small periodic gauge transformations and are central in modern analyses of gauge theories on small circles and tori \cite{Hosotani83,AitkenChermanPoppitzYaffe2017,Akamatsu2022}. Motivated by this, we keep only the transverse components and set
\begin{align}
A_t=A_x=0,
\qquad
A=A_{\mathfrak y}(t,x)\,d\mathfrak y + A_{\mathfrak z}(t,x)\,d\mathfrak z.
\end{align}
In this sector $F-G$ carries no $(t,x)$-dependence, so $\partial_t(F-G)=\partial_x(F-G)=0$. Consequently, the gauge-invariant combination $\partial_\mu(F-G)+2A_\mu$ has $(t,x)$ components equal to $2A_t$ and $2A_x$; setting $A_t=A_x=0$ is then consistent with the equations of motion and amounts to working in a representative where only the transverse holonomies are retained.

The specific form of the Ansatz is fixed by how the $U(1)$ acts on the Euler angles. Since $U\to e^{\lambda t_3}Ue^{-\lambda t_3}$ shifts
\begin{align}
F\to F+\lambda,
\qquad
G\to G-\lambda,
\qquad
H\to H,
\end{align}
the combination $F-G$ transforms as $F-G\to F-G+2\lambda$. It follows that the derivative combination
\begin{align}
\partial_\mu(F-G)+2A_\mu
\end{align}
is gauge-invariant under $A_\mu\to A_\mu-\partial_\mu\lambda$, and it is precisely the object that controls how the covariant derivative reorganizes in the Euler-angle variables. We introduce the Euler combinations
\begin{align}
u\equiv F+G,
\qquad
v\equiv F-G,
\end{align}
so that $u$ is gauge-invariant while $v$ shifts as $v\to v+2\lambda$, and we name the corresponding gauge-invariant building block
\begin{align}
X_\mu\equiv \partial_\mu v+2A_\mu.
\end{align}
In particular, one can check directly from $R_\mu=U^{-1}D_\mu U$ and the Euler parametrization that the quadratic sigma model invariant reorganizes diagonally into the form
\begin{align}
-\frac12\,\text{Tr}(R_\mu R^\mu)
= \partial_\mu H\partial^\mu H
+\cos^{2}\!H\,\partial_\mu u\,\partial^\mu u
+\sin^{2}\!H\,X_\mu X^\mu,
\end{align}
which makes explicit that the $U(1)$ gauge field couples only through $X_\mu$, i.e.\ only to the $v$-channel.

With the chiral choice $F=p\,\mathfrak z$ and $G=p\,\mathfrak y$, one has
\begin{align}
\partial_{\mathfrak y}(F-G)=-p,
\qquad
\partial_{\mathfrak z}(F-G)=+p.
\end{align}
If one kept the transverse zero-modes without compensating these rigid gradients, the constants $\pm p$ would enter explicitly through the gauge-invariant combinations $\partial_{\mathfrak y}v+2A_{\mathfrak y}$ and $\partial_{\mathfrak z}v+2A_{\mathfrak z}$, so local densities would retain a hard imprint of the transverse winding. In the minimal interacting truncation adopted in the main text, it is therefore natural to let the transverse gauge zero-modes absorb this rigid data, leaving a single dynamical \(1+1\) degree of freedom. Introducing $\psi(t,x)$, we choose
\begin{align}
2A_{\mathfrak y}=p-L\psi(t,x),
\qquad
2A_{\mathfrak z}=-(p-L\psi(t,x)),
\end{align}
so that
\begin{align}
\partial_{\mathfrak y}(F-G)+2A_{\mathfrak y}=-L\psi(t,x),
\qquad
\partial_{\mathfrak z}(F-G)+2A_{\mathfrak z}=+L\psi(t,x).
\end{align}
Equivalently, in terms of $v$ and $X_\mu$ this choice enforces the minimal symmetric conditions
\begin{align}
X_{\mathfrak y}=\partial_{\mathfrak y}v+2A_{\mathfrak y}=-L\psi(t,x),
\qquad
X_{\mathfrak z}=\partial_{\mathfrak z}v+2A_{\mathfrak z}=+L\psi(t,x),
\end{align}
so that the $v$-channel contribution to $X_\mu X^\mu$ becomes purely $(t,x)$-dependent. Indeed, using ${g^{\mathfrak y\mathfrak y}=g^{\mathfrak z\mathfrak z}=1/L^{2}}$ one has
\begin{align}
X_\mu X^\mu \supset \frac{X_{\mathfrak y}^{2}}{L^{2}}+\frac{X_{\mathfrak z}^{2}}{L^{2}}
=\frac{(-L\psi)^{2}}{L^{2}}+\frac{(+L\psi)^{2}}{L^{2}}
=2\psi^{2},
\end{align}
so the sigma model sector produces the characteristic reduced interaction $\propto \psi^{2}\sin^{2}\!H$ once the Ansatz is imposed.

It is worth stressing that the transverse zero-mode reduction on the two compact cycles would generically leave two independent \(1+1\) scalar zero-modes, carried by $A_{\mathfrak y}(t,x)$ and $A_{\mathfrak z}(t,x)$, corresponding to the two Wilson-line (holonomy) modes of the transverse Maxwell components on $T^2$. A convenient way to display this is to introduce the linear combinations
\begin{align}
A_{+}\equiv \frac{1}{2}\big(A_{\mathfrak y}+A_{\mathfrak z}\big),
\qquad
A_{-}\equiv \frac{1}{2}\big(A_{\mathfrak z}-A_{\mathfrak y}\big),
\end{align}
so that the transverse one-form can be written as $A_{\mathfrak y}\,d\mathfrak y+A_{\mathfrak z}\,d\mathfrak z=A_{+}\,(d\mathfrak y+d\mathfrak z)+A_{-}\,(d\mathfrak z-d\mathfrak y)$. In the present topological sector the gauged Euler combination $v\equiv F-G$ winds as $v\propto(\mathfrak z-\mathfrak y)$, so the gauge-invariant building blocks $\partial_{\mathfrak y}v+2A_{\mathfrak y}$ and $\partial_{\mathfrak z}v+2A_{\mathfrak z}$ single out the $d\mathfrak z-d\mathfrak y$ channel as the one that is needed to absorb the rigid $\pm p$ data. The choice above implements precisely the minimal consistent sector $A_{+}=0$ (equivalently $A_{\mathfrak y}+A_{\mathfrak z}=0$), leaving a single dynamical transverse mode, here parametrized by $\psi(t,x)$. Retaining $A_{+}$ would add a second \(1+1\) scalar degree of freedom; it can be consistently frozen by choosing its holonomy to be constant (taken here to be zero), in which case it is not sourced by the winding-driven dynamics and remains decoupled in the strict zero-mode truncation. Any mixing of $A_{+}$ back into the interacting $(H,\psi)$ sector arises only beyond that truncation (e.g.\ from non-zero KK modes or higher-dimension operators) and is suppressed by the EFT cutoff set by the transverse KK scale $1/L$ (and, if smaller, by the intrinsic UV scale of the microscopic theory).

In components $(t,x,\mathfrak y,\mathfrak z)$ this is
\begin{align}
A_\mu=\Big(0,\,0,\,\tfrac12\big(p-L\psi(t,x)\big),\, -\tfrac12\big(p-L\psi(t,x)\big)\Big),
\end{align}
which matches the gauge Ansatz in \eqref{AnsatzGauge}.

It is also useful to record explicitly how the Maxwell term reduces and where the kinetic term for $\psi$ comes from. Considering the abelian field strength $F_{\mu\nu}\equiv \partial_\mu A_\nu-\partial_\nu A_\mu$, the truncation $A_t=A_x=0$ implies that the only potentially nonzero components are those with one index in $(t,x)$ and the other in $(\mathfrak y,\mathfrak z)$. With the Ansatz above one finds
\begin{align}
F_{t\mathfrak y}=\partial_t A_{\mathfrak y}=-\frac{L}{2}\,\partial_t\psi,
\qquad
F_{x\mathfrak y}=\partial_x A_{\mathfrak y}=-\frac{L}{2}\,\partial_x\psi,
\end{align}
and similarly
\begin{align}
F_{t\mathfrak z}=\partial_t A_{\mathfrak z}=+\frac{L}{2}\,\partial_t\psi,
\qquad
F_{x\mathfrak z}=\partial_x A_{\mathfrak z}=+\frac{L}{2}\,\partial_x\psi.
\end{align}
Using the cavity metric, in particular $g^{\mathfrak y\mathfrak y}=g^{\mathfrak z\mathfrak z}=1/L^{2}$, one sees that $F_{\mu\nu}F^{\mu\nu}$ is proportional to $-(\partial_t\psi)^2+(\partial_x\psi)^2$, so the Maxwell Lagrangian density $-\frac{1}{4e^2} F_{\mu\nu}F^{\mu\nu}$ generates the standard \(1+1\) kinetic structure for $\psi$ (up to the overall normalization fixed in the main text when passing to canonically normalized fields). If the additional zero-mode $A_{+}$ were retained, it would similarly appear as a second \(1+1\) scalar with its own Maxwell-induced kinetic term.

With the combined choices for $U$ and $A_\mu$, gauge-invariant local densities built from ${R_\mu=U^{-1}D_\mu U}$ and from the field strength become independent of $(\mathfrak y,\mathfrak z)$, so the field equations close consistently on the reduced fields $\{H(t,x),\psi(t,x)\}$. Meanwhile, the discrete label $p$ remains as global data tied to the compact transverse cycles, and the rigid transverse gradients coexist with the longitudinal profile $H(t,x)$, allowing nontrivial topological/baryonic structure in a finite-volume setting of the kind explored in finite-volume Skyrmion constructions \cite{CanforaDiMauroKurkovNaddeo2015}.

\newpage

\bibliographystyle{JHEP}
\bibliography{ChiralPhoton}

\end{document}